\documentclass[a4paper,11pt]{article}
\pdfoutput=1 

\usepackage{jheppub} 

\usepackage[T1]{fontenc} 
\usepackage[utf8]{inputenc}
\usepackage{amsmath}
\usepackage{amsfonts}
\usepackage{amssymb} 
\usepackage{graphicx}
\usepackage{siunitx}

\usepackage[dvipsnames]{xcolor}
\usepackage[colorinlistoftodos]{todonotes}
\usepackage{commath}
\usepackage{xspace}
\usepackage{url}
\newcommand{\wimpsim}{\texttt{WimpSim}\xspace}
\newcommand{\wimpann}{\texttt{WimpAnn}\xspace}
\newcommand{\wimpevent}{\texttt{WimpEvent}\xspace}
\newcommand{\solarcrnu}{\texttt{solar\_crnu}\xspace}
\newcommand{\nusigma}{\texttt{nusigma}\xspace}
\newcommand{\pythia}{\texttt{Pythia}\xspace}
\newcommand{\darksusy}{\texttt{DarkSUSY}\xspace}
\newcommand{\mceq}{\texttt{MCEq}\xspace}
\newcommand{\ds}{\texttt{DarkSUSY}\xspace}
\newcommand{\sibyll}{\texttt{SIBYLL}\xspace}

\newcommand{\numu}{{\nu_\mu}}
\newcommand{\nue}{{\nu_e}}

\newcommand{\sanu}{SA$\nu$\xspace}

\title{\boldmath Neutrinos from cosmic ray interactions in the Sun}

\author[a]{J.~Edsj{\"o},} 
\author[a,*]{J.~Elevant\note[*]{Corresponding authors},}
\author[b]{R.~Enberg,} 
\author[a,*]{C.~Niblaeus}

\affiliation[a]{Oskar Klein Centre for Cosmoparticle Physics,
Department of Physics, Stockholm University, SE-10691 Stockholm, Sweden}
\affiliation[b]{Department of Physics and Astronomy, Uppsala University, Box 516, SE-75120 Uppsala, Sweden}

\emailAdd{edsjo@fysik.su.se}
\emailAdd{jessica.elevant@fysik.su.se}
\emailAdd{rikard.enberg@physics.uu.se}
\emailAdd{carl.niblaeus@fysik.su.se}

\abstract{Cosmic rays hitting the solar atmosphere generate neutrinos that interact and oscillate in the Sun and oscillate on the way to Earth. These neutrinos could potentially be detected with neutrino telescopes and will be a background for searches for neutrinos from dark matter annihilation in the Sun. We calculate the flux of neutrinos from these cosmic ray interactions in the Sun and also investigate the interactions near a detector on Earth that give rise to muons. We compare this background with both regular Earth-atmospheric neutrinos and signals from dark matter annihilation in the Sun. Our calculation is performed with an event-based Monte Carlo approach that should be suitable as a simulation tool for experimental collaborations. Our program package is released publicly along with this paper.}

\begin{document}

\maketitle
\flushbottom

\section{Introduction}

Just as in the Earth's atmosphere, cosmic rays (CRs) hitting the solar atmosphere create a cascade that eventually generates neutrinos. The aim of this study is to estimate the size of this flux including interactions and oscillations. 

If we look towards the Sun, the Sun will block some CRs and we expect to get a lower Earth atmospheric neutrino signal from the direction of the Sun. On the other hand these blocked CRs will produce neutrinos in the Sun instead. Naively we would expect these to be of similar magnitude, but the typical density where neutrinos are produced is lower in the solar atmosphere and neutrinos also experience interactions and oscillations on the way to Earth. The stronger magnetic fields in the Sun will also affect both the CRs on their way to the Sun and the cascade development in the Sun. Hence, we expect the neutrino fluxes from the Sun to be different compared to the fluxes from the Earth's atmosphere. This motivates studying these fluxes in more detail.

These neutrinos from the solar atmosphere can teach us (if observed) about both the primary cosmic rays, the solar density in the outskirts of the Sun, the solar magnetic fields and possibly about neutrino oscillations.
They will also be a background for other searches for high energy neutrinos from the Sun, most notably for searches for dark matter via dark matter annihilations in the Sun that can give rise to neutrinos \cite{Press:1985ug,Silk:1985ax}. We will compare our results with the signal from WIMP (Weakly Interacting Massive Particle) annihilations.

These solar atmospheric neutrinos, \sanu, have been studied in the past in e.g. refs.~\cite{Moskalenko:1991jr,Moskalenko:1991hm,Seckel:1991ffa,Moskalenko:1993ke,Ingelman:1996mj,Hettlage:1999zr,Fogli:2006jk}. In the first four of these studies, neutrino oscillations were not included, in the fifth one they were included partially, and in the last two they used the production fluxes in the atmosphere calculated by Ingelman \& Thunman in ref.~\cite{Ingelman:1996mj} (hereafter denoted IT96) and estimated the effect of oscillations. Our focus here is to improve on these earlier studies by calculating the complete process from cosmic ray interactions in the Sun through interactions and oscillations to a flux of neutrinos and neutrino-induced leptons at a neutrino telescope. Our improvements mainly come from using newer cosmic ray models, hadronic interaction models in the solar atmosphere, an improved solar density model and a better simulation of the neutrino interaction and oscillation effects in the Sun and oscillations on the way to a detector at Earth.  Adding the effects of magnetic fields presents a considerable challenge however and we will not include those effects in this study. Instead we restrict ourselves to neutrino energies above 50 GeV where the effects from the solar magnetic fields are expected to be small.

Our approach is event-based using Monte Carlo techniques. For the CR interactions in the solar atmosphere we use the publicly available tool \mceq \cite{Fedynitch:2015zma} which we have modified to work for the Sun instead of the Earth's atmosphere. For the interactions and oscillations in the Sun, oscillations between the Sun and the Earth and interactions in the Earth close to the detector we use an updated version of the publicly available simulation package \wimpsim \cite{Blennow:2007tw,Edsjo:2007ws}. Our additions to that code to include the \sanu{}s are also made publicly available \cite{Edsjo:2017ws}.

\section{Cosmic ray interactions and neutrino production in the Sun}

\subsection{Production mechanism and cascade}

\begin{figure}
\begin{center}
\includegraphics[scale=1]{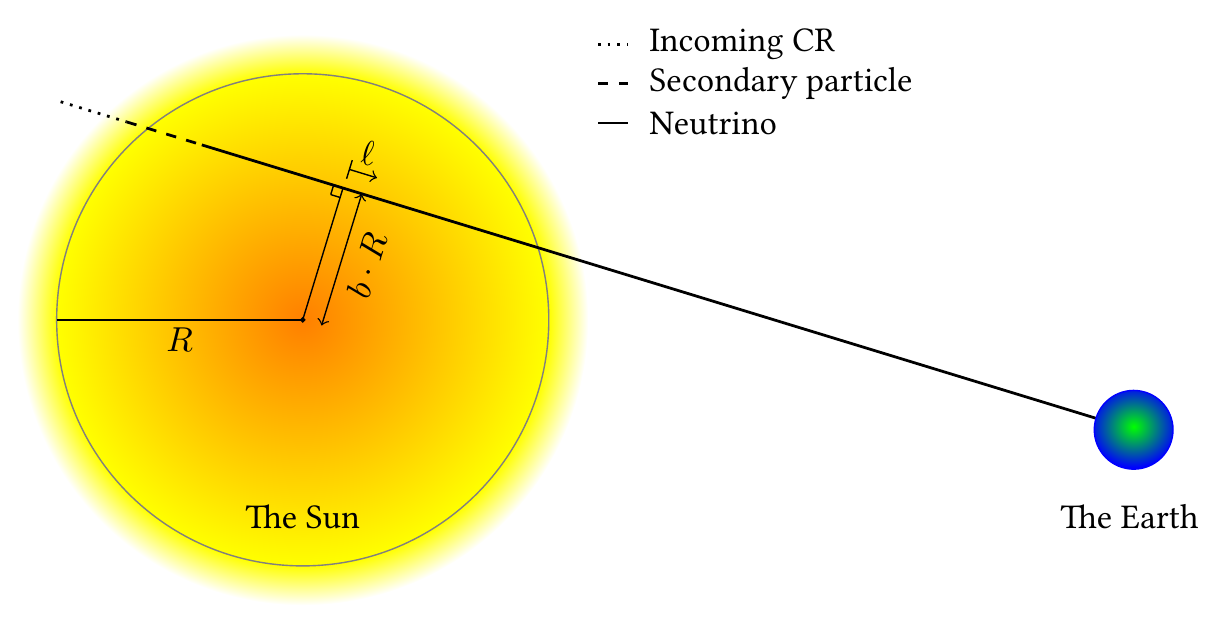}
\caption{A schematic geometry showing how the particles travel through the Sun. Incoming CRs interact with the Sun creating secondary particles which decay into/interact creating neutrinos. The length parameter $\ell$ is defined to be $0$ at the point closest to the centre of the Sun and follows the trajectory of the incoming CR at an impact parameter $b$ and continues all the way to the Earth.}
\label{fig:geometry}
\end{center}
\end{figure}

When CRs impinge on the Sun they collide with nuclei in the Sun, resulting in the development of hadronic cascades. These cascades contain a plethora of hadrons and leptons that decay and interact further, resulting in a flux of neutrinos. The neutrinos are mainly produced from leptonic decays of pions and kaons. This is described by a chain of decays that begins with the initial hadronic collision producing $\pi^+$ ($\pi^-$) or $K^+$ ($K^-$) that subsequently decay into a $\mu^+$ ($\mu^-$) and $\nu_\mu$ ($\bar{\nu}_\mu$). The $\mu^+$ ($\mu^-$) further decay into $e^+ \nu_e \bar{\nu}_\mu$ ($e^- \bar{\nu}_e \nu_\mu$). The resulting flux of neutrinos is called the \emph{conventional} flux and has an approximate flavour ratio of 
$(\nu_e+\bar{\nu}_e):(\nu_{\mu}+\bar{\nu}_{\mu}):(\nu_{\tau}+\bar{\nu}_{\tau})=1:2:0.$

Apart from the conventional neutrino flux there is a contribution to the neutrino flux called the \emph{prompt} flux. This is induced by  decays of charmed mesons such as $D^0$ and $D^{\pm}$. Due to the higher masses and shorter lifetimes of these mesons, the prompt flux is mainly important at higher energies, where the conventional flux falls off faster with energy due to energy losses of the long-lived pions and kaons.
We will here assume that the effects of the solar magnetic fields is negligible and that the cascade is developing in the direction of the primary CR particle. This is a good approximation at higher energies, but will be worse for lower energies, see Section \ref{sec:crmagn} below for more details.

The development of the cascade is described by a set of coupled differential equations that describe how the flux of each particle type depends on the atmospheric slant depth $X$. The slant depth is for a specific trajectory  from $\ell_0$ to $\ell$ given by the integral of the atmospheric density $\rho$ along the path:
\begin{equation}
X(\ell) = \int_{\ell_0}^{\ell} \rho(\ell') \ d \ell'
\end{equation}
where $\ell$ is a variable tracing the trajectory of the particle and $\rho(\ell)$ is the density at the point $\ell$. In terms of $X$ the cascade equation for the flux of a particle type $i$ at energy $E$ is written

\begin{subequations}
\begin{align}
\begin{split}
\label{eq:casceqa}
	\frac{d\Phi_i (E)}{dX}= & -\frac{\Phi_i(E)}{\lambda_{i,\mathrm{int}}(E)}
\end{split} \\
\begin{split}
\label{eq:casceqb}
	& -\frac{\Phi_i(E)}{\lambda_{i,\mathrm{dec}}(X,E)}
\end{split} \\
\begin{split}
\label{eq:casceqc}
	& + \sum_j \int_{E'>E} \frac{\Phi_j(E')}{\lambda_{j,\mathrm{int}}(E')} \frac{d n_{j(E')\rightarrow i(E)}}{ d E} \ d E'
\end{split} \\
\begin{split}
\label{eq:casceqd}
	& +  \sum_j \int_{E'>E} \frac{\Phi_j(E')}{\lambda_{j,\mathrm{dec}}(X,E')} \frac{d n_{j(E')\rightarrow i(E)}}{dE} \ d E'.
\end{split}
\end{align}
\end{subequations}

The first two terms are sink terms, while the last two terms are source terms. They represent an decrease and increase respectively in the flux of particle $i$.

The first term, \eqref{eq:casceqa}, describes interactions where particles of type $i$ are lost and is governed by the interaction length 
\begin{equation}
\lambda_{i,\mathrm{int}}(E)= \frac{\rho}{n\sigma_{i\mathrm{-atm}}^{\mathrm{inel}}(E)} \approx \frac{\langle m_{\mathrm{atm}} \rangle }{\sigma_{i\mathrm{-atm}}^{\mathrm{inel}}(E)}
\end{equation}
which, in the approximation that $\rho(X)/n(X)\approx \langle m_{\mathrm{atm}}\rangle$, i.e. that the atmospheric composition is independent of depth ($n$ is the atmospheric number density), is the atmospheric mass average divided by the (weakly energy dependent) inelastic cross section for interactions of particle $i$ with the atmospheric particles. It is in this approximation independent of the slant depth $X$. The average atmospheric mass is given by $\langle m_{\text{atm}} \rangle=m_p\langle A\rangle$ where $m_p$ is the proton mass and $\langle A\rangle = 1.27$ the average mass number for the nuclei in the solar atmosphere, assuming a composition of 28 \% helium and 72 \% hydrogen. We use the inelastic cross section for particle $i$ on protons, $\sigma_{i\,p}^{\rm inel}$, for the interaction lengths.
This means that we approximate the cross section $\sigma_{i\mathrm{-atm}}^{\rm inel} \simeq \sigma_{i\,p}^{\rm inel}$, but adjust the mass average to the actual conditions in the Sun. We estimate that this approximation is good to within about 10\%. The second term, \eqref{eq:casceqb}, describes particle loss due to decays and contains the decay length 
\begin{equation}
\lambda_{i,\mathrm{dec}}(X,E)=c\tau_i \rho(X) \frac{E}{m_i}
\end{equation}
where $\tau_i$ is the life-time of particle $i$ and $E/m_i=\gamma$ represents time dilation. The decay length depends explicitly on the energy and implicitly on the slant depth through the density $\rho(X)$.

The two last terms in the cascade equation, \eqref{eq:casceqc} and \eqref{eq:casceqd}, represent an increase of the flux of particle $i$ due to interactions  and decays containing the particle in the final state. They contain integrals over the fluxes of other particles $j$ and the corresponding interaction/decay lengths multiplied by the yields $dn_{j(E')\rightarrow i(E)}/dE$ --- the number of $i$ particles at energy $E$ coming from the interaction/decay of particle $j$ with energy $E'$.

The coupled system of cascade equations can be solved in a number of ways, ranging from purely numerical to semi-analytical methods that introduce the spectrum-weighted $Z$-moments. In this case the equations can be solved separately in the limits of high and low energies, where the interaction term and the decay term dominate, respectively, and the solution is then an interpolation between these two solutions. There is then a critical energy above which the flux falls off faster by one power of the energy.

\subsection{The \mceq code and modifications}

In this paper we use the code \mceq \cite{Fedynitch:2015zma,mceqonline} to obtain the neutrino fluxes. In this code the cascade equations are formulated in matrix form, with the fluxes for all particle types in a column vector and the different interaction/decay lengths and regeneration terms as matrix elements, and solved by utilising methods from linear algebra.  

Originally, the \mceq code was meant to treat CR cascades in the Earth's atmosphere, but we have modified it to make it possible to obtain results for cascades in the solar atmosphere. We have (i) included muon energy loss as described in section~\ref{sec:muon-decay}, (ii) changed to a solar geometry, (iii) used solar density profiles as described in section~\ref{sec:densityprofiles}, and (iv) used cross sections and particle yields appropriate for a solar atmospheric environment. 

In \mceq the CR flux is converted into fluxes of neutrons and protons with energies divided equally among the nucleons. These are then followed along with all other included particle types in the cascade equations. Tabulated  cross sections and yields must be provided to \mceq. Several event generators are available to provide these, we use the cross sections and yields for hadronic and baryonic projectiles on protons obtained with the event generator \sibyll \cite{Fletcher:1994bd,Ahn:2011wt} version 2.3 \cite{Riehn:2015oba,Engel:2015dxa}.

The prompt flux is often modelled separately from the conventional flux. In this paper we use the default option in \mceq, i.e.\ the charm production model of \sibyll~2.3~\cite{Engel:2015dxa}. This is a phenomenological model, in contrast to the modern calculations of the prompt flux in the atmosphere~\cite{Bhattacharya:2015jpa,Garzelli:2015psa,Gauld:2015kvh,Bhattacharya:2016jce} that use state-of-the-art perturbative QCD. The \sibyll prompt flux is larger than the perturbative fluxes, and may be close to the IceCube upper limit on the prompt flux~\cite{Radel:2015rsz,Bhattacharya:2016jce}. However, as the prompt flux in Earth's atmosphere is only relevant for neutrino energies above 10$^6$ GeV, we do not expect it to be important for solar neutrinos in the energy range we are interested in.

\subsection{The Sun}
\label{sec:densityprofiles}

\begin{figure}
\centering
\includegraphics[width=0.49 \textwidth]{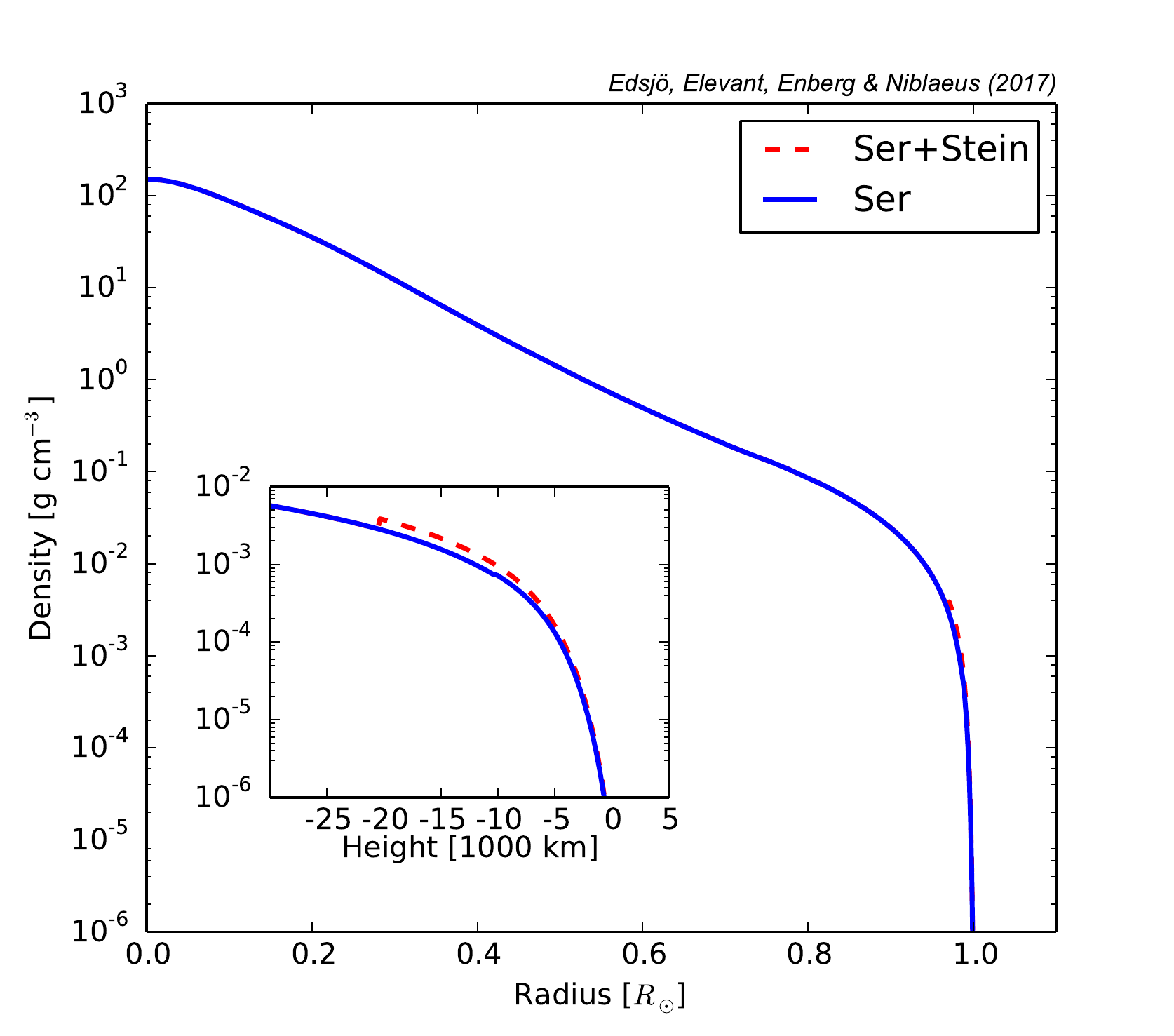}
\caption{Shown here are the radial density profiles we have considered.  The solid blue curve is a Standard Solar Model by Serenelli et al. \cite{Serenelli:2009yc} while the dashed red line comes from a Serenelli et al.  model out to $\sim$\SI{20000}{\kilo\meter} below the solar surface, where we change to the density from a magnetohydrodynamic simulation by Stein et al. \cite{Stein1998ApJ}. For both curves we use the exponential fit to data in ref.~\cite{Vernazza1981} used by Ingelman et al.~\cite{Ingelman:1996mj} outside the solar surface. The density makes a discontinuous jump where the model changes which is unphysical, we have made the choice to use the densities at face value rather than making a possibly erroneous matching of the curves at points of model change.}
\label{fig:densityprof}
\end{figure}

For the radial density profile, describing the distribution of mass in the Sun as a function of the distance from the centre, we use the Standard Solar Models (SSMs) for the interior of the Sun. The SSMs mainly concern the inner parts of the Sun and close to the surface we need to provide additional models. We use two benchmark models. For the first benchmark, \emph{Ser+GS98} we use the Serenelli \cite{Serenelli:2009yc} model combined with Grevesse \& Sauval \cite{Grevesse:1998bj}. For the second model, \emph{Ser+Stein}, we use ref.~\cite{Serenelli:2009yc} for the interior and the density resulting from a numerical simulation by Stein et al. \cite{Stein1998ApJ}\footnote{As obtained from Bob Stein's webpage: \url{http://steinr.pa.msu.edu/~bob/data.html}.} closer to the surface (above $\sim$\SI{20000}{\kilo\meter}). The latter is a magnetohydrodynamical simulation of convection and magnetic fields near the solar surface. We use the average density in the vertical direction. Outside the solar surface, we use, for both benchmark models, the same density profile as in the IT96 study which is an exponential fit to data in ref.~\cite{Vernazza1981}. The two resulting density profiles we use in our calculations are plotted in figure \ref{fig:densityprof}, but in practice, the choice of density model does not significantly affect the results. We have also checked that the Serenelli model is very similar to Model S by Christensen-Dalsgaard et al.\ \cite{ChristensenDalsgaard:1996ap}. The density curve jumps discontinuously at the points where the model changes. This is unphysical, however we have made the choice to keep the values as is rather than making a guess for how to match the curves. The jump does not affect our results in a significant way.

The magnetic field of the Sun is the source of the interplanetary magnetic field that extends throughout the Solar System. In the outer parts of the Sun, in the corona and chromosphere, the magnetic field has a complicated structure and our knowledge to a large part relies on numerical modelling and approximations  \cite{Wiegelmann2014, Mackay:2012ww, Riley2006} using observations of the magnetic field in the photosphere as boundary conditions. The magnetic flux is unevenly distributed on the solar surface and is concentrated in flux tubes that extend outwards from the surface, and there is a toroidal magnetic field around the equator. The magnetic field in the flux tubes at the surface can be in excess of $10^3$ G. This will affect the cascade development that we will discuss in the next subsection.

In order to estimate energy losses of charged particles in the cascade we need to know the magnetic field inside the Sun in the regions where the neutrinos are generated. The complicated structure of the solar magnetic field is thought to have its origin at the base of the convective zone, at roughly $0.7R_\odot$, and it is believed that the magnetic fields there are at least $10^4$ G or up to $10^5$ G in flux tubes~\cite{solarmagnetic}. Because the \sanu are produced closer to the surface, we do not need to know the magnetic fields deep into the Sun.

\subsection{Cosmic rays, the solar magnetic field and cascade developments}
\label{sec:crmagn}

CRs that reach the solar atmosphere will be affected by the solar magnetic field. The effect will be largest for the lower energy particles --- at higher energies the particles are rigid enough that the magnetic effects can be neglected. The effect on the CRs up until their first interaction in the Sun has been studied in Ref.~\cite{Seckel:1991ffa}, where the neutrino energy $E_m$ below which magnetic effects are important is estimated to be $E_m \sim \SI{200}{\giga\electronvolt}$. CRs leading to neutrinos of energies below $E_m$ are in their analysis assumed to partially prevented from reaching the Sun, hence the effect in Ref.~\cite{Seckel:1991ffa} is to lower the neutrino flux below $E_m$.

The magnetic field will also affect the motion of charged particles in the cascades, causing their direction to change. Hence, the cascade development will in general not only be in the direction of the primary particle. One effect of this would be that the cascade does not develop as far down into the atmosphere as without a magnetic field, another effect is the potential that the particles in the cascades are mirrored.

A gamma-ray flux from the Sun has been observed by the Fermi-LAT \cite{Abdo:2011xn,Ng:2015gya}. The flux is about ten times larger than that previously expected \cite{Seckel:1991ffa} and furthermore has a time-dependence anti-correlated with the solar activity \cite{Ng:2015gya}. The gamma ray observations can potentially aid in determining the effect of the magnetic fields on the low energy neutrino flux since part of the gamma ray flux is produced by the $\pi^0$ decay in the same hadronic cascades that give the neutrino flux. In fact, as shown in ref.~\cite{Seckel:1991ffa}, the gamma ray flux will be enhanced by the mirroring of incoming CRs in the magnetic fields. These effects on the cascade development should in principle also affect the neutrino flux in a similar fashion. The fact that the gamma ray observations are higher than expected indicate that there can be an enhancement in the neutrino flux.

In the current analysis we neglect the effects of magnetic fields. Given the above discussion, this should be valid for neutrinos of higher energies. We follow ref.~\cite{Seckel:1991ffa} and judge that the magnetic effects is relevant for neutrino energies below $E_m\sim \SI{200}{\giga\electronvolt}$ with increasing importance the lower the energy is. As the effects of the magnetic fields are expected to be larger the lower the neutrino energy is we will in this study show results a bit below 200 GeV and will cut our fluxes at 50 GeV. In future work we will attempt to estimate the effects from the magnetic field.


\subsection{Energy loss mechanisms}
\label{sec:energy-loss}

The particles created in the cascade may lose energy in several ways. The main mechanism is hadronic interactions, as indicated above. In the generation of neutrino fluxes in Earth's atmosphere, this is the only relevant mechanism, and as such, it is the only mechanism implemented in \mceq by default. In neutrino production in astrophysical sources, however, hadronic interactions are not the main energy loss mechanism, but instead radiative cooling, i.e.\ synchrotron radiation and inverse Compton scattering, are more important due to the very large magnetic fields and photon densities.
The Sun has a significant magnetic field and number density of photons, so in principle it may be possible for particles to lose energy through these radiative processes. However, as we shall see, these two mechanisms for energy loss can be safely neglected in the Sun and the energy loss implemented in \mceq is sufficient. Below we give a simple estimate of the relative importance of the three energy loss mechanisms, and since the radiative processes are not important we do not include them in our calculations. It would, however, be straightforward to include them if necessary, provided that a sufficiently good knowledge of the magnetic field was available. Including radiative cooling would only be relevant for ultra-high energies, however, where the fluxes are very low.

As we shall see below, most of the neutrino production occurs between $r=0.99R_\odot$ and $r=1R_\odot$. For definiteness, we choose the density and temperature below at $r=0.995R_\odot$ for illustration, and we choose a magnetic field strength of $10^5$ G, but the conclusions do not depend on this choice.

The attenuation length for synchrotron losses for protons of energy $E_p$ in a magnetic field $B$ is 
\begin{equation}
\lambda_{p,\text{synchro}} = \frac{6\pi m_p^4 c^4}{\sigma_{T}m_e^2E_p B^2},
\end{equation}
where $\sigma_T$ is the Thomson cross section for electrons. This should be rescaled by $(m_M/m_p)^4$ for a meson of mass $m_M$. Note that the attenuation length is in units of cm. 

With a density at $r=0.995 R_\odot$ of $5\times 10^{-5}$ g/cm$^3$, the hadronic interaction length for protons, $\lambda_{p,\text{int}}$, is on the order of $10^5$ cm with a logarithmic dependence on energy. The synchrotron length $\lambda_{p,\text{synchro}}$ falls with energy and magnetic field strength, but for all energies of interest it is orders of magnitude larger than $\lambda_{p,\text{int}}$. For pions, the hadronic length is roughly the same as for protons, while the synchrotron length is rescaled by a factor $5\times 10^{-4}$. This is not enough to make them comparable, except for the extreme choice of $B=10^5$ G, where they become equal for very large energies $E_\pi\sim 10^{11}$ GeV. Note that if a perhaps more realistic value of $B\sim 10^3$ G is used, they instead become equal at $10^{15}$ GeV. We therefore conclude that synchrotron losses are completely negligible in the energy range we consider.

For inverse Compton scattering on a photon energy density $U_\gamma$ the attenuation length is
\begin{equation}
\lambda_{p,\text{IC}} = \frac{3m_p^4c^4}{4 \sigma_{T} m_e^2 E_p U_\gamma},
\end{equation}
where $U_\gamma\propto T^4$ is the photon energy density as obtained from the Stefan-Boltzmann law for temperature $T$. The solar models discussed above give the temperature as a function of the radial distance from the centre, with a temperature at $r=0.995 R_\odot$ around $2.5\times 10^4$ K. This gives the estimate that $\lambda_{p,\text{IC}}$ is $10^5$ times larger than $\lambda_{p,\text{synchro}}$. Inverse Compton scattering can therefore also be neglected.

It should be noted that the above estimates use a conservative value of $B=10^5$ G, and if the true value is smaller, synchrotron cooling becomes even more irrelevant.

In the next section we discuss the energy loss of muons in particular, which is one of our additions to \mceq. Muons are also affected by the radiative processes discussed in this section, but since the muon mass is quite close to the pion mass, they will have roughly the same attenuation lengths such that we may again neglect radiative energy loss.


\subsection{Muon energy loss and decay}
\label{sec:muon-decay}

Muons lose energy, which in \mceq is modelled in a continuous fashion according to
\begin{equation}
\frac{dE}{dX}=-(\alpha + \beta E)
\end{equation}
where $E$ is the muon energy, $X$ the slant depth and $\alpha$ and $\beta$ parameters that depend on the material that the muons propagate through. As muons decay and produce neutrinos it is important to include the energy losses.

As in ref.~\cite{Seckel:1991ffa} we model the outer parts of the Sun as consisting of 72~\% hydrogen and 28~\% helium which gives us the values \num{7.0e-3} GeV cm$^2$ g$^{-1}$ and \num{1.8e-6} cm$^2$ g$^{-1}$ for $\alpha$ and $\beta$ respectively. We have produced tables of energy losses as a function of energy and use these in \mceq.

As the muons lose energy, their decay length decreases, and for most impact parameters they all decay within the Sun and produce neutrinos. However, for very high impact parameters, there is not enough solar material to propagate through, so some muons will decay after the Sun.  \mceq only calculates the decays inside the Sun and to include the neutrinos from the decay of these remaining muons we have added an extension where we let the muons decay manually and add the resulting neutrinos to the neutrinos produced in \mceq.

\subsection{Cosmic ray models}
\label{sec:cr-models}

The all-particle CR spectrum is well modelled by a power-law with a single slope of around $-2.7$ up to energies of about $10^6$ GeV, at which point the spectrum becomes steeper (the so called \emph{knee}). At energies around $10^9$ GeV the spectrum changes again at the \emph{ankle}, signalling a probable transition to an extragalactic source of CRs since the energy is then too high for the CRs to be contained by the galactic magnetic field. There are various options for the parametrization of the CR spectrum in \mceq. The models differ in their assumptions on what types of CR populations make up the spectrum. Typically one assumes galactic and extragalactic components in the spectrum, with the extragalactic populations being relevant mainly from the knee and upwards. 

In our analysis, we use two different models: the Hillas-Gaisser 3-generation model (denoted \emph{H3a}) model \cite{Gaisser:2011cc} and the Gaisser-Stanev-Tilav 4-generation model (here denoted \emph{GST 4-gen}) model \cite{Gaisser:2013bla}. In the \emph{H3a} model, three different populations of CRs are assumed, one extragalactic component that starts to contribute to the spectrum at the ankle and two galactic components below the ankle. The \emph{GST 4-gen} model assumes four populations, two of galactic origin and two extragalactic, the fourth one consisting of purely protons included to make the CR composition less heavy at the highest energies.


\begin{figure}
\centering
\includegraphics[width=0.32\textwidth]{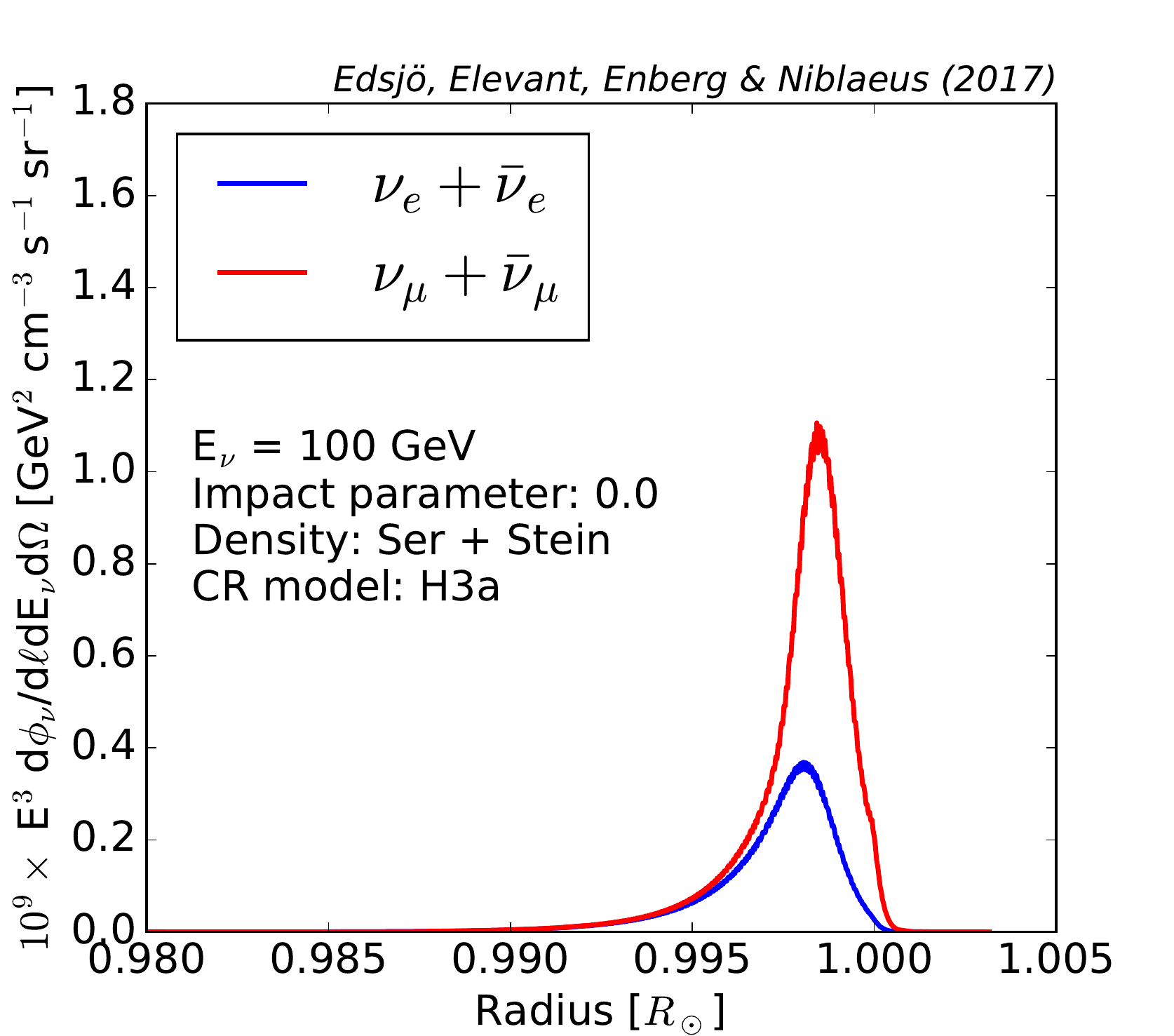}
\includegraphics[width=0.32\textwidth]{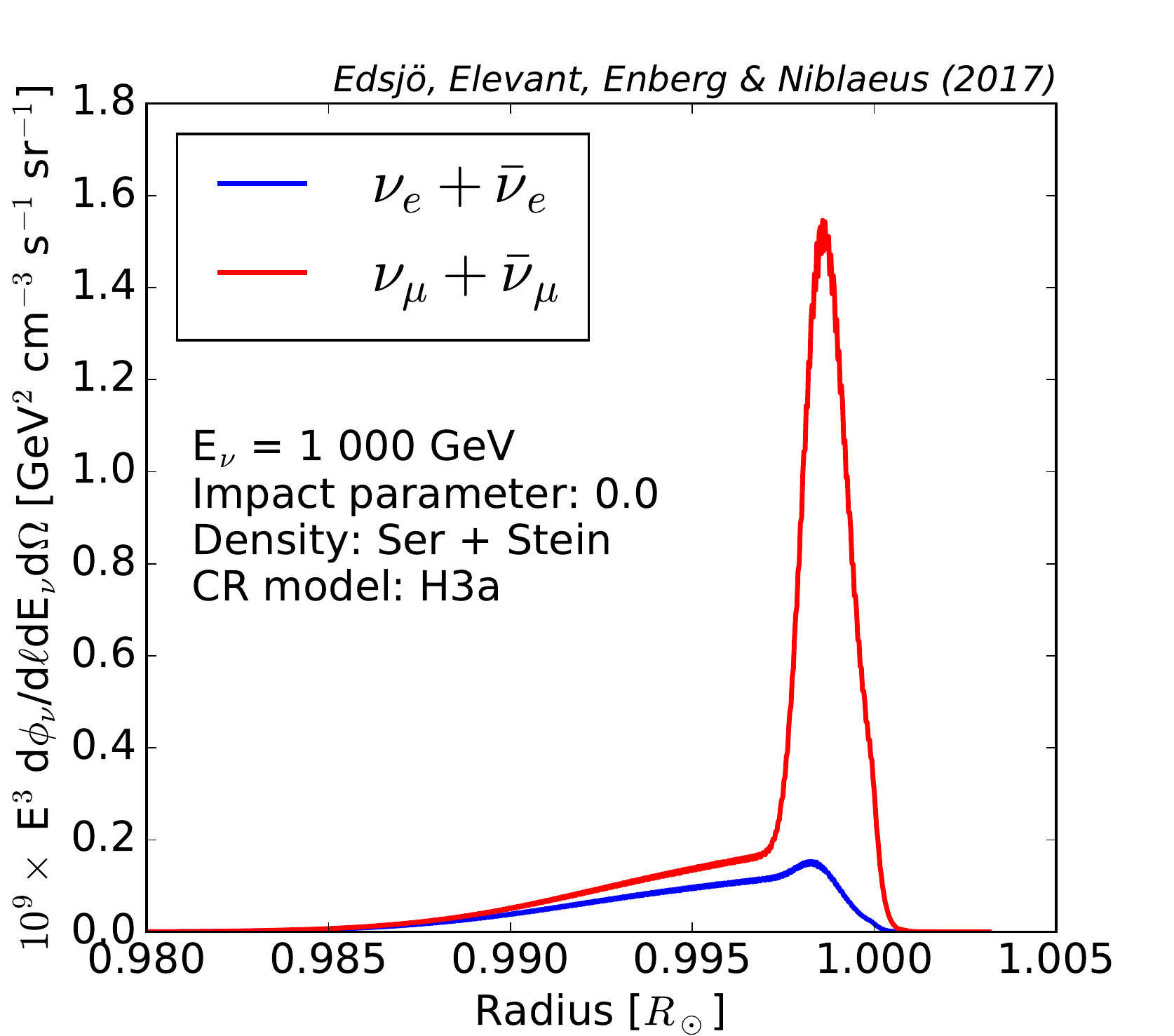}
\includegraphics[width=0.32\textwidth]{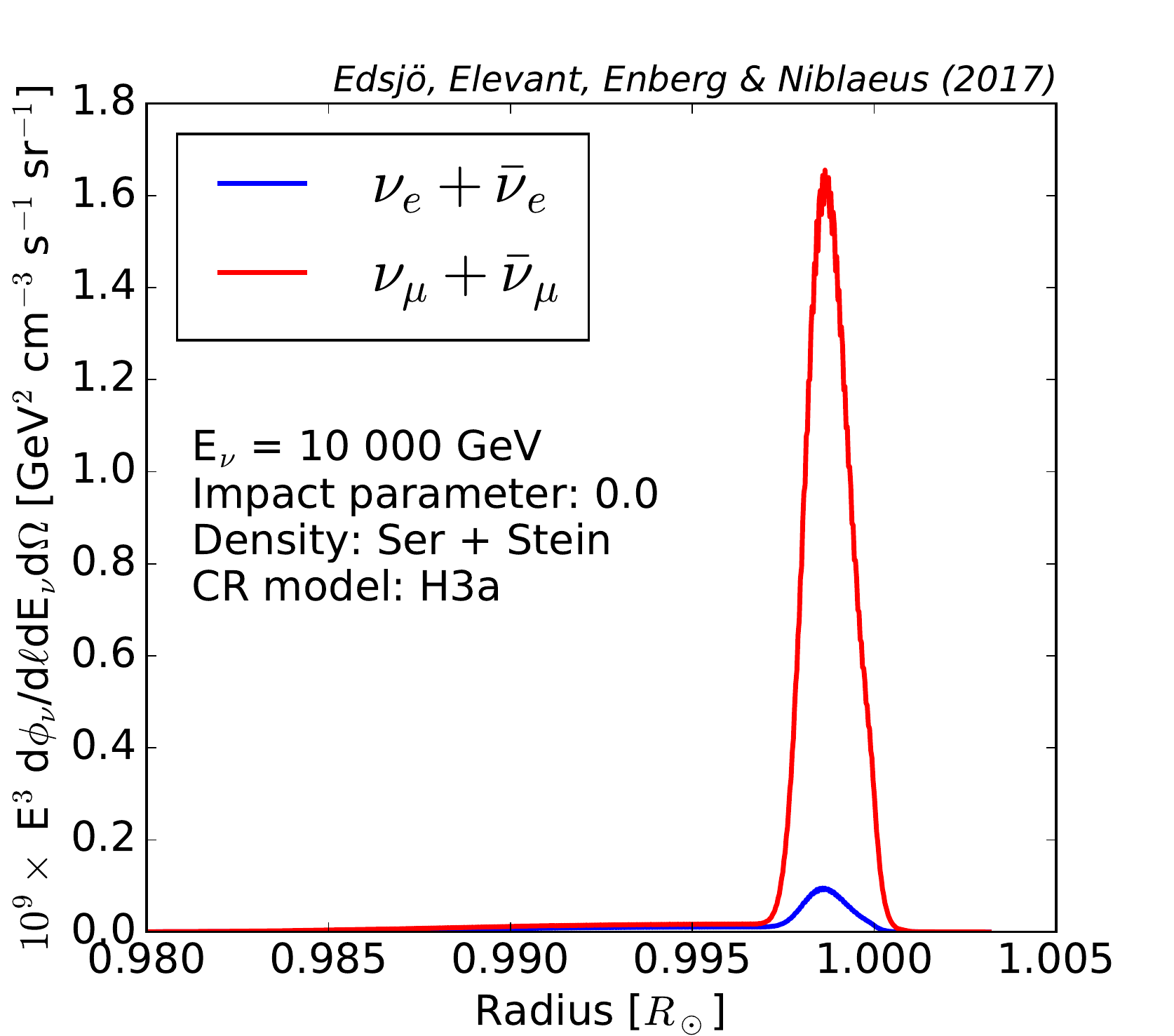}
\caption{Differential neutrino flux with respect to length travelled for an impact parameter $b = 0$ and three different energies (100 GeV, 1 000 GeV and 10 000 GeV). Note that the neutrinos are dominantly produced just beneath the surface of the Sun.}
\label{fig:dphidl}
\end{figure}

\subsection{Neutrino fluxes at production}

In the outer parts of the Sun, cascade particles (mainly pions, kaons and muons) will decay into neutrinos. In figure \ref{fig:dphidl}, we show the flux differentiated with respect to length travelled to show where neutrinos are produced. We show this for an impact parameter $b = 0$ and three different energies. Note, that most of the neutrinos are produced directly beneath the surface of the Sun. The position of the production peak depends on impact parameter, but most of the neutrinos are produced between 0.99$R_\odot$ and 1.0$R_\odot$.

At a depth sufficiently large that all cascade particles that produce neutrinos will have decayed, no more neutrinos are produced by the CR interactions with the solar matter. We call these fluxes our production fluxes. We also include neutrinos arising from muon decay outside the Sun (as described in section \ref{sec:muon-decay}) in these production fluxes. In section \ref{sec:nuprod} we will come back to these fluxes after having discussed how we perform the event generation in the next section.

\section{Neutrino interactions and oscillations in the Sun and interactions at the detector on Earth} 
\label{sec:intosc}

From the \mceq calculations we have the neutrino production fluxes as a function of impact parameters and length travelled in the solar atmosphere. These neutrinos will then propagate through the Sun undergoing interactions and oscillations.  
For the interactions we include deep inelastic scattering via both neutral current (NC) and charged current (CC) interactions. NC will degrade the energy of the neutrino and we include these lower energy neutrinos in our calculation. In CC interactions we will produce a charged lepton. For electrons and muons these will not give rise to new neutrinos (muons are considered stopped before they decay), but for $\tau$ leptons, these will decay and produce new lower energy neutrinos, which we include in our calculation. 

In this section we will go through how we in our event-based Monte Carlo framework use the \mceq production fluxes as input and draw events from these distributions and take care of neutrino interactions and oscillations. For this we use the publicly available \wimpsim code \cite{Blennow:2007tw,Edsjo:2007ws}. The \wimpsim code is created to take care of neutrino interactions and oscillations for neutrinos arising from WIMP dark matter annihilations in the Sun and the Earth. In this study, we have modified \wimpsim to include also the \sanu{}s.

\subsection{Event generation in \wimpsim}\label{sec:eventgen}

For a fixed value of the impact parameter $b$, \mceq provides the fluxes of neutrinos produced in the cascades as function of flavour, energy and path length $\ell$ (or depth $X$) travelled. For definitions of path length $\ell$ and impact parameter $b$, see figure~\ref{fig:geometry}. We differentiate these to obtain the differential fluxes with respect to path length $d\Phi_{\nu_{\alpha}}/d\ell$ and read the differential fluxes into \wimpsim for 14 values of the impact parameter ranging from 0 to 1.002 (we have picked the values of $b$ to make sure we sample the distribution well, especially close to the limb, $b\simeq 1$). The fluxes are then interpolated linearly in $b$, $\ell$ and $\log{E}$. We have chosen our set of impact parameters to ensure small interpolation errors.

We generate events uniformly distributed in the impact parameter $b$ and assign each event a weight to properly include the $b$ dependence on the solid angle. We then sample neutrino events with a value of energy from the integrated flux $\Phi$ using acceptance-rejection sampling. Each event is also given an $\ell$-value by using acceptance-rejection sampling on the differential flux distribution $d\Phi_{\nu_{\alpha}}/d\ell$. In practice we draw events from the distributions summed over flavour and particle/antiparticle type and assign flavour and particle/antiparticle type from their relative probabilities.

Both energy and $\ell$ span over many orders of magnitude, with a rather steep decline. In order to obtain reasonably good statistics at all orders of magnitude, we used rejection sampling under a curve that is as close to the sampling distribution as possible. It takes about a minute to generate 1 million events.

In this way we obtain neutrino events with energies, impact parameter and flavour ratios distributed in accordance with the \mceq fluxes, and points of creation distributed according to the differential flux so that most neutrinos come from points mostly around an energy dependent distance of the order of a few thousands of kilometres below the solar surface. Each event is assigned a weight that ensures that the sum of all events will give the flux in units of $1/$(cm$^2$ s).

\subsection{Neutrino interactions}

\begin{figure}
\centering
\includegraphics[width=0.49\textwidth]{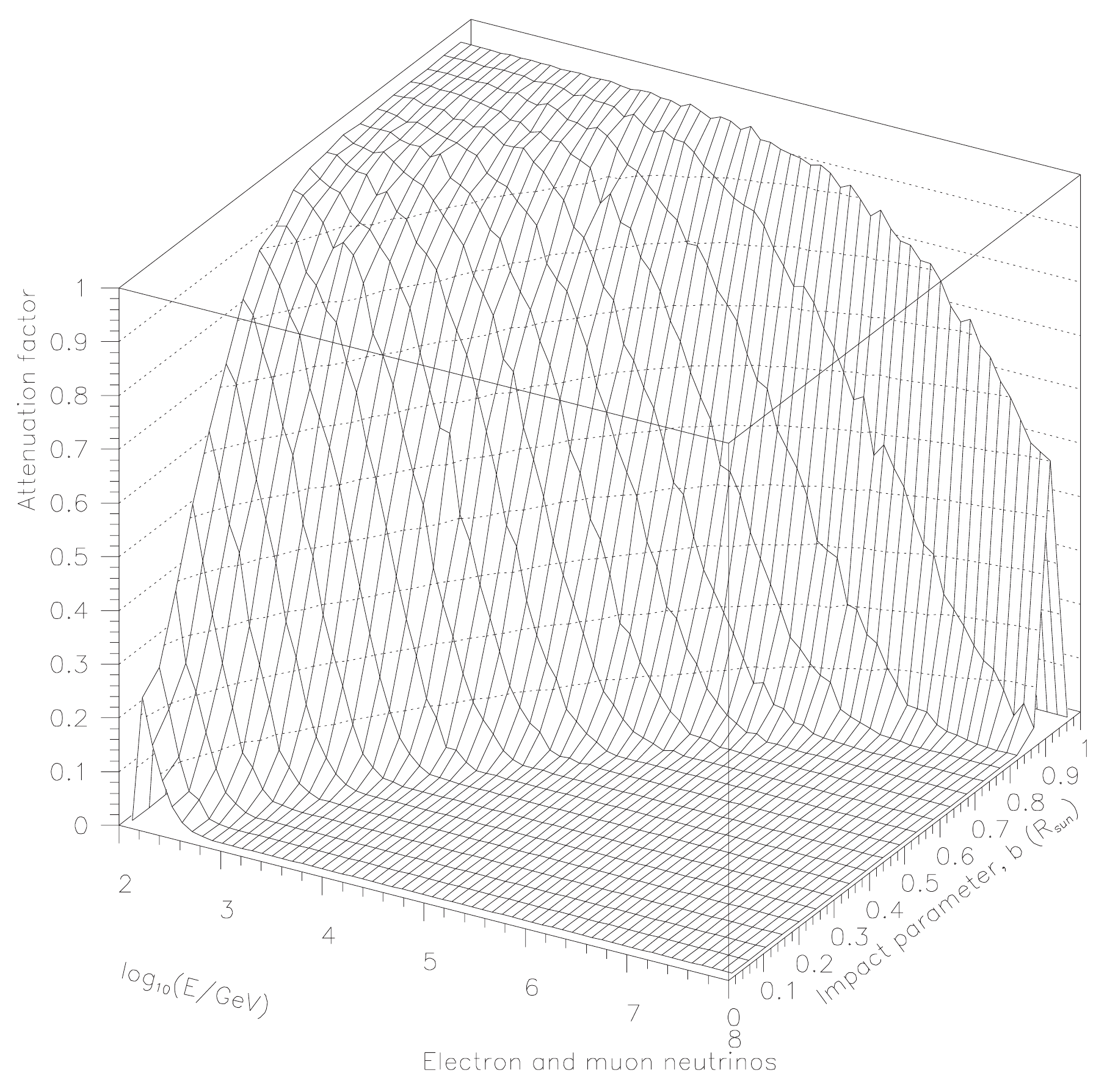}
\includegraphics[width=0.49\textwidth]{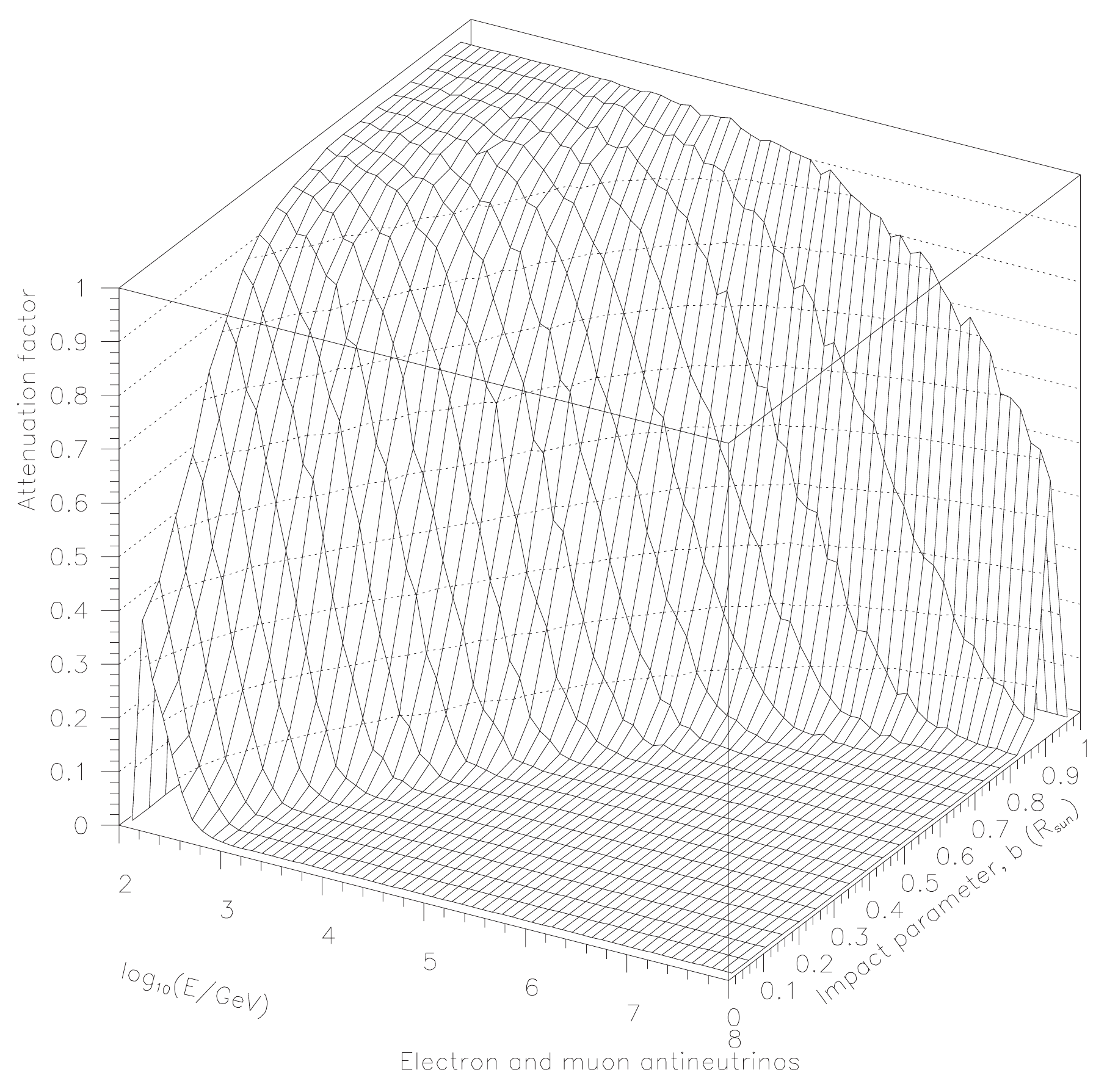}
\caption{An illustration of the attenuation of $\nu_e$ and $\nu_\mu$ (left) and $\bar{\nu}_e$ and $\bar{\nu}_\mu$ (right) through CC and NC interactions in the Sun. The figure shows the fraction of neutrinos that survive passage through the Sun for different energies and impact parameters. In this plot we have for simplicity assumed that both NC and CC interactions will absorb the neutrinos, whereas in the real calculation we include the lower energy neutrinos from NC interactions and from $\tau$ decay. This figure is dedicated to the memory of PAW \cite{paw}.}
\label{fig:attenuation}
\end{figure}

When the generated neutrinos propagate through the Sun on their way to the Earth they will interact and undergo flavour oscillations. As we focus on high energy neutrinos, we consider interactions through deep inelastic scattering on the nuclei in the solar interior, where the neutrino exchanges energy with a parton inside a nucleon through weak vector boson exchange. The interactions include NC and CC interactions, proceeding through the exchange of a $Z$ and $W$ boson respectively. In NC interactions, the neutrino is kept, but with a lower energy. In CC interactions we produce a charged lepton, which for electrons and muons will not give rise to new neutrinos (muons are considered stopped before they decay in the region where interactions are significant). If a $\tau$ lepton is produced it will decay and produce new lower energy neutrinos. The neutrino-nucleon interactions are simulated with \nusigma \cite{Edsjo:2007ns} using the CTEQ6-DIS parton distribution functions \cite{Kretzer:2003it} and $\tau$ lepton decay we simulate
with \pythia 6.4.26 \cite{Sjostrand:2006za}.

The interactions are simulated from the point in the Sun where the neutrino is created, and onward using the same density profile as for CR interactions. We use the composition in the Serenelli SSM \cite{Serenelli:2009yc} to include interactions on the correct amount of protons and neutrons throughout the Sun (i.e.\ we do not assume an isoscalar target). We link to \ds where these density profiles and the composition in the Sun are easily obtained \cite{Gondolo:2004sc}. In case a charged current interaction that creates a $\tau$ lepton takes place, we simulate the tau decay and inject the produced neutrinos at this location and continue simulating interactions and decay through the Sun.

In figure~\ref{fig:attenuation} we illustrate how important the interactions are by calculating the attenuation factors in a simple setup. We have here assumed that both NC and CC interactions will absorb the neutrinos, whereas in the real calculation we include the lower energy neutrinos from NC interactions and from $\tau$ decay. This figure shows the attenuation for electron and muon neutrinos and antineutrinos.\footnote{For tau neutrinos, the results would be similar, but the CC cross section is slightly lower for tau neutrinos due to the $\tau$ lepton mass.} We can see from these figures that for low impact parameters ($b=0$) we lose essentially all neutrinos above $10^3$ GeV, whereas close to the solar surface ($b=1$) the neutrinos are not very much affected by interactions below around $10^5$ GeV. These figures can be compared to the earlier calculation IT96~\cite{Ingelman:1996mj}. Qualitatively our results are quite similar, but we do get a higher suppression, especially for low impact parameters. Most likely this is due to that we use an updated solar density model and neutrino-nucleon cross sections compared to the IT96 study. We will come back to the effect the interactions have on our \sanu{}s in section \ref{sec:nuprod}.


\subsection{Neutrino oscillations}
\label{sec:neutrinoOscillations}

Neutrino oscillations are included in \wimpsim using a full three-flavour numerical evolution code. It steps through the Sun taking both vacuum oscillations, oscillations from matter effects and interactions into account (as described in the previous subsection). For the matter effects, we use the Serenelli SSM \cite{Serenelli:2009yc} for the electron density in the Sun. We will perform our calculations for three different sets of oscillation parameters: one without oscillations and one each for the best fit normal and inverted mass ordering scenarios. The best fit values we take from refs.\ \cite{Esteban:2016qun,nufitonline}.
These sets of values are listed in table \ref{tab:oscparams}.

\begin{table}[h!]
\caption{Table of the three sets of oscillation parameters we use. For normal ordering and inverted ordering the best fit values are from refs.~\cite{Esteban:2016qun,nufitonline}.}
\begin{center}
\begin{tabular}{| c | r r r r r r |}
\hline
& $\theta_{12} (^\circ)$ & $\theta_{23} (^\circ)$ & $\theta_{13} (^\circ)$ & $\delta_{CP} (^\circ)$ & $\Delta m^2_{21}$ (eV$^2$) & $\Delta m^2_{31}$ (eV$^2$) \\
\hline
No osc. & 0 & 0 & 0 & 0 & $0^\text{*}$ & $0^\text{*}$ \\
Normal ordering & 33.56 & 41.6 & 8.46 & 261 & $7.50\cdot 10^{-5}$ & $+2.524\cdot 10^{-3}$ \\
Inverted ordering & 33.56 & 50.0 & 8.49 & 277 & $7.50\cdot 10^{-5}$ & $-2.439\cdot 10^{-3}$ \\
\hline
\multicolumn{7}{l}{\footnotesize{$^{*}$In the code, we set the mass squared differences to non-zero values to avoid numerical problems.}}\\
\end{tabular}
\end{center}
\label{tab:oscparams}
\end{table}

The probabilities for neutrino oscillations are oscillating functions with amplitudes determined by the neutrino mixing angles $\theta_{ij}$ and oscillation lengths that depend on the neutrino energy and the squared mass differences $\Delta m_{ij}^2$. Thus we get three oscillation lengths: $\lambda_{21}$, $\lambda_{31}$ and $\lambda_{32}$. Since $\lvert\Delta m_{31}^2\rvert \approx \lvert\Delta m_{32}^2\rvert$, $\lambda_{31}\approx \lambda_{32}$ and we effectively have two different oscillation lengths. In the vacuum approximation, where matter oscillation effects are ignored, these are approximately given by
\begin{subequations}
\begin{align}
\begin{split}
\label{eq:osc-length1}
\lambda_{21} & \approx \SI{3.3e6}{\kilo\meter} \left( \frac{E}{\SI{100}{\giga\electronvolt}} \right) \left(\frac{\SI{7.5e-5}{\electronvolt\squared}}{\Delta m_{21}^2}\right) 
\end{split} \\
\begin{split}
\label{eq:osc-length2}
\lambda_{3i} &\approx \SI{9.9e4}{\kilo\meter} \left( \frac{E}{\SI{100}{\giga\electronvolt}} \right) \left( \frac{\SI{2.5e-3}{\electronvolt\squared}}{\lvert \Delta m_{3i}^2 \rvert} \right), \ i=1,2,
\end{split}
\end{align}
\end{subequations}
i.e. $\lambda_{21} \simeq R_{\odot}$ for energies above \SI{100}{\giga\electronvolt} while all oscillation lengths are small compared to the Sun-Earth distance for energies around \SI{100}{\giga \electronvolt}. We can divide the effects of oscillations for each oscillation length in three regions: i) low energies, where oscillations become decoherent and the oscillating part can be averaged over ii) intermediate energies, where  oscillation effects can be seen in the neutrino fluxes and iii) high energies, where the  oscillation lengths are long compared to the Sun-Earth distance so that neutrinos do not have time to oscillate. 

Although we include the full treatment of matter effects on the neutrino oscillations, these  effects are  not significant for the energies we are considering, as also shown earlier in ref.~\cite{Fogli:2006jk}. The oscillation effects on the neutrino fluxes at Earth come dominantly from vacuum oscillations.

\subsection{\wimpsim running}

Our new solar atmospheric neutrino code \solarcrnu is added to the existing \wimpsim software and made publicly available \cite{Edsjo:2007ws}. For details about the code, we refer the reader to Appendix \ref{app:wimpsim} and the code webpage. We will also provide the result files used in this paper on the \wimpsim web page \cite{Edsjo:2007ws}.  

In \wimpsim we can simulate the events at a particular detector on Earth and we will assume that the detector is IceCube \cite{Achterberg:2006md} located at latitude $-90^\circ$, with the detector medium being ice. We will further assume that the data taking window is between the vernal and autumn equinoxes (i.e.\ during the austral winter where the Sun is below the horizon and hence the atmospheric muon background is lower). We will in this study focus on the summary fluxes which will contain the fluxes we are interested in, time-averaged over the six months of the austral winter. For the time-averaging, the eccentricity of the Earth's orbit is included which will cause some of the oscillation patterns to be washed out. Our assumption on the detector medium being ice will only affect our neutrino-induced muon fluxes and using a different detector location or medium surrounding the detector will typically change the fluxes by less than 5\%.

\section{Results: neutrino fluxes and production neutrino fluxes and neutrino-induced muon fluxes at production and at the Earth}

We are now ready to show some resulting fluxes, where we will focus on fluxes at the detector on Earth. We have calculated the neutrino fluxes for two cosmic ray models, \emph{H4a} and \emph{GST 4-gen}, two density profiles, \emph{Ser+Stein} and \emph{Ser+GS98} and three neutrino oscillation scenarios, no oscillations, normal ordering and inverted ordering, i.e.\ in total 12 different combinations. For each combination we have generated \num{2.5e8} neutrinos. As a default, we will show results for the cosmic ray model \emph{H3a} and the density profile \emph{Ser+Stein}, but will also investigate some of the dependencies we have on different input parameters.

\subsection{Neutrino fluxes at production and after passage through the Sun}
\label{sec:nuprod}

\begin{figure}
\centering
\includegraphics[width=0.49\textwidth]{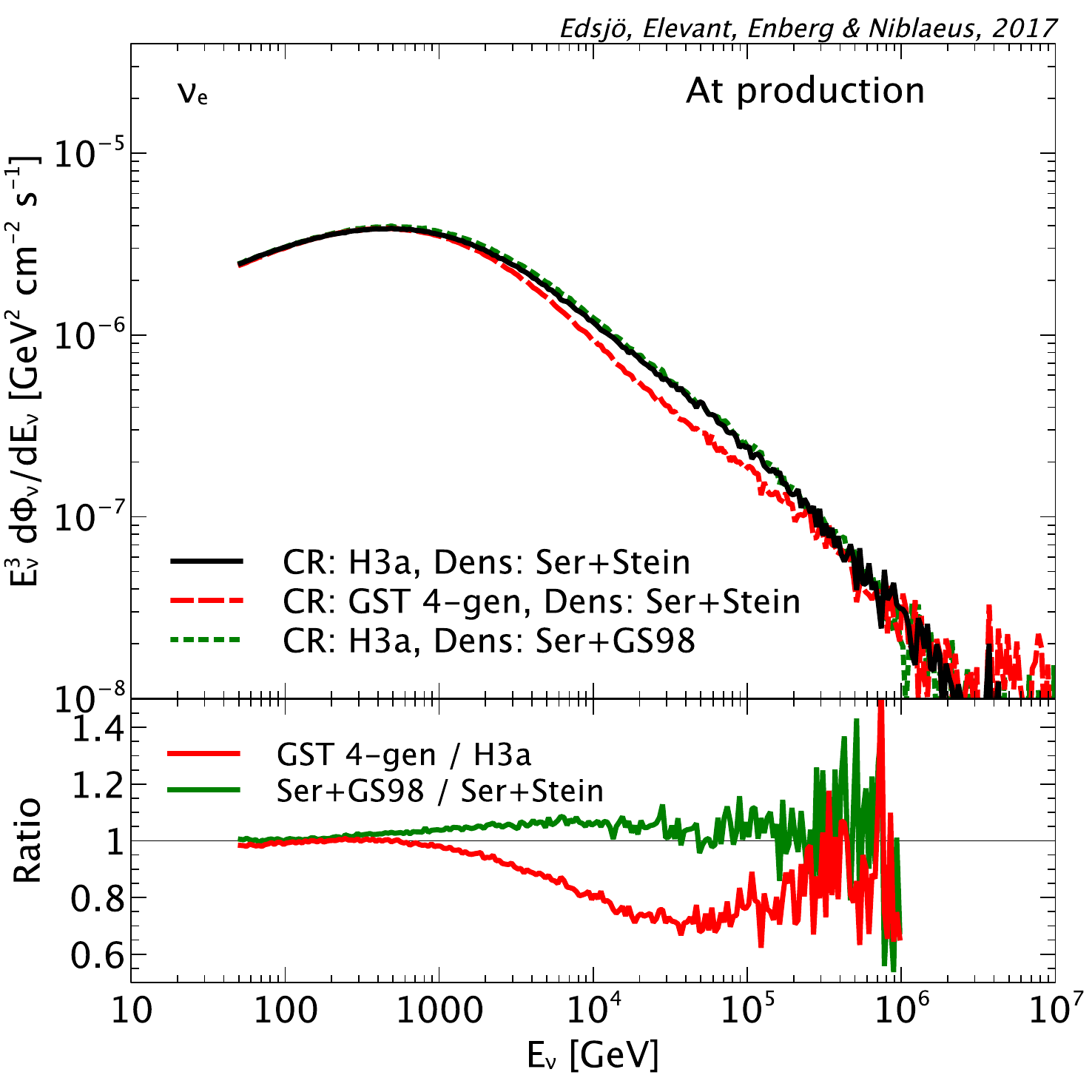}
\includegraphics[width=0.49\textwidth]{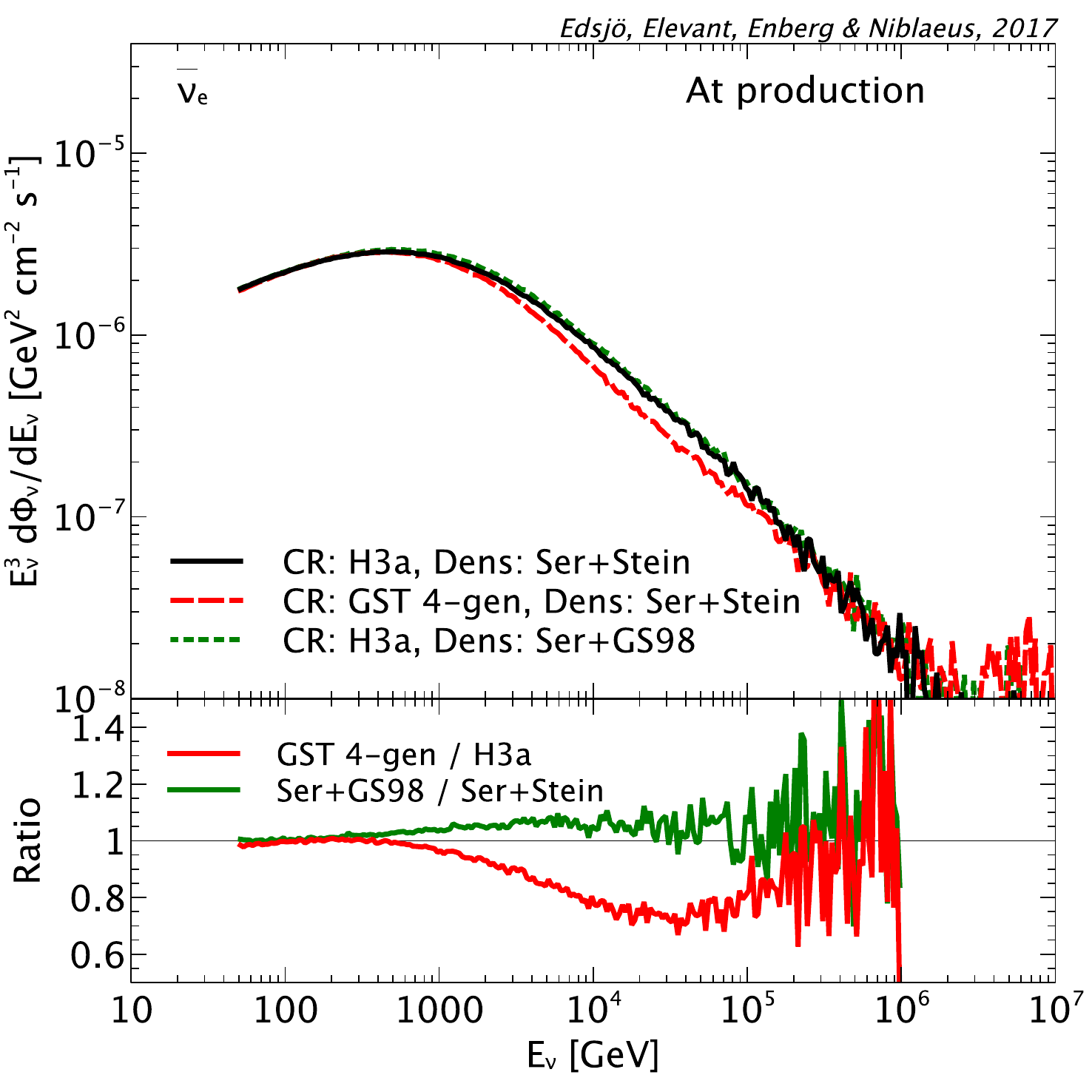}
\includegraphics[width=0.49\textwidth]{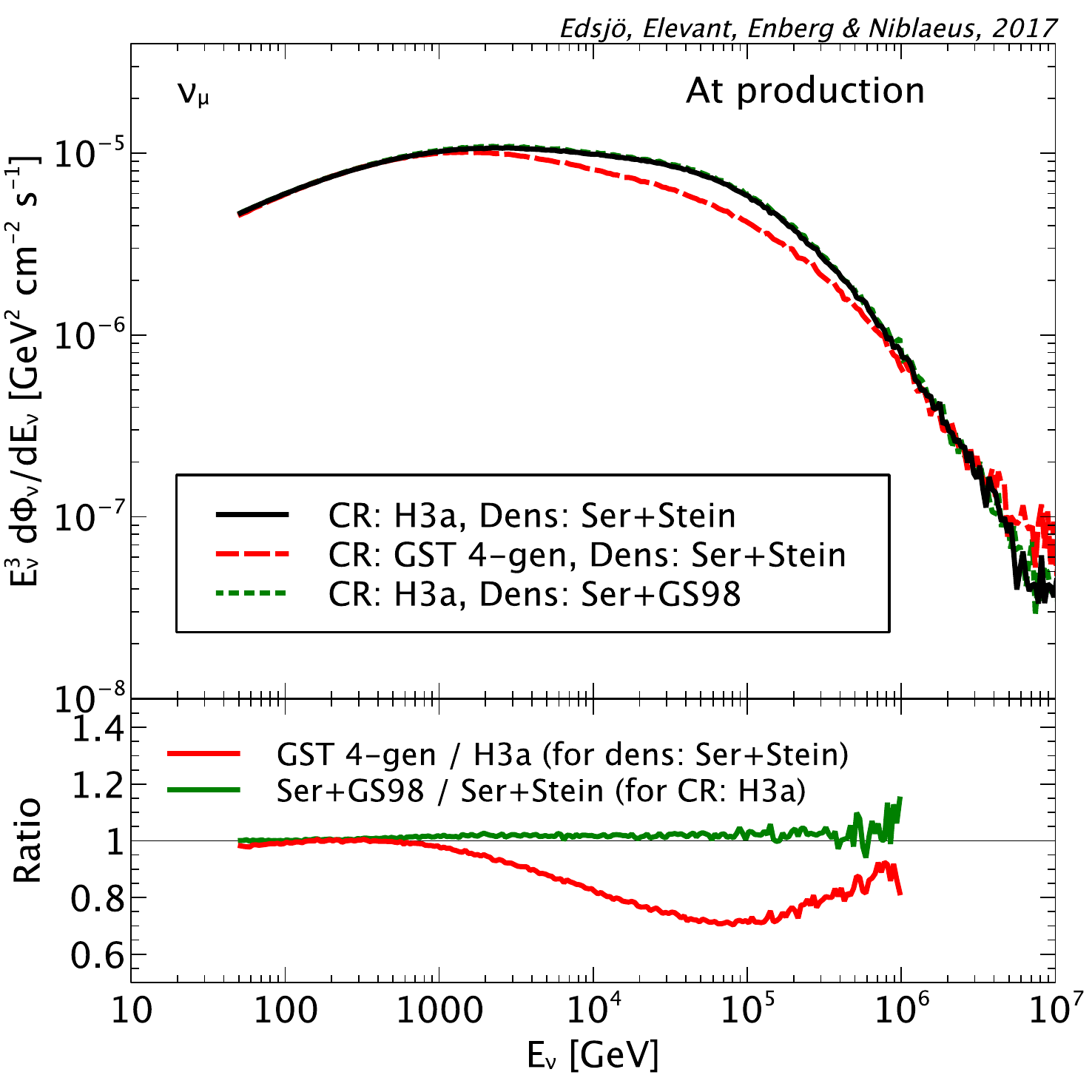}
\includegraphics[width=0.49\textwidth]{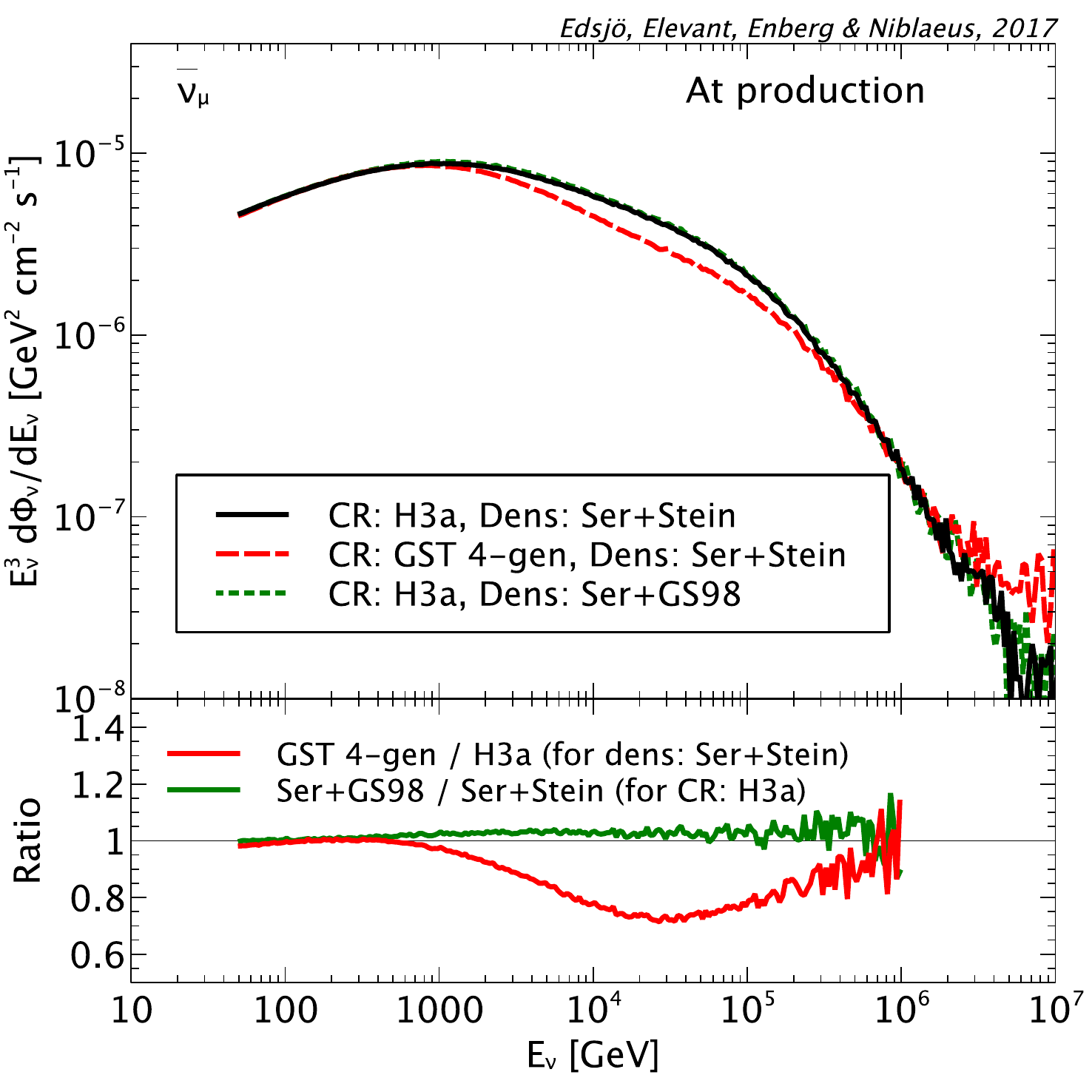}
\caption{In the upper row we show the production fluxes of $\nu_e$ (left) and $\bar{\nu}_e$ (right) respectively. In the lower row, we show the same but for muon neutrinos and anti-neutrinos. We show the production fluxes for our default cosmic ray model \emph{H3a} and density profile \emph{Ser+Stein}, but also compare the fluxes from a different cosmic ray model \emph{GST 4-gen} and density profile \emph{Ser+GS98}. In the lower part of the figures we show the ratio between the fluxes for different cosmic ray models and density profiles.}
\label{fig:prodfluxes}
\end{figure}

Our production fluxes for $\nue$ and $\numu$ are shown in figure~\ref{fig:prodfluxes} for our different choices of CR models and density profiles. The fluxes are as seen from Earth, integrated over the solid angle of the sun. We see that the production fluxes can differ by up to 30\% at higher energies depending on which CR model we use. At lower energies (below around 100 GeV), the differences are very small though. Our two different density models give very similar results with the differences typically being less than 5\%. 
As our two density models only differ significantly at the solar surface, we expect to see the biggest differences from the two models in the production fluxes, the interactions and oscillations will be very similar for our two models.

\begin{figure}%
\centering
\includegraphics[width=0.32\textwidth]{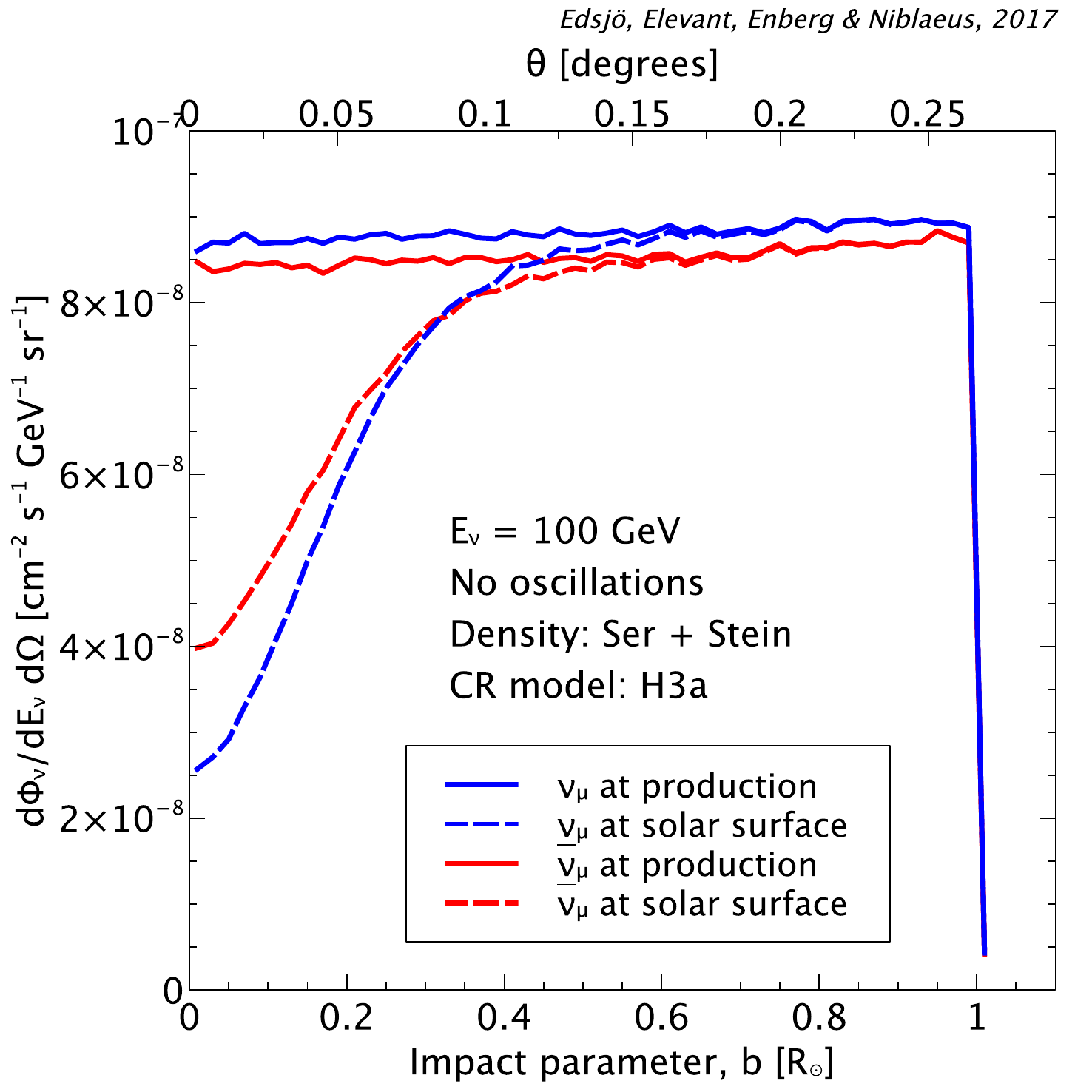}
\includegraphics[width=0.32\textwidth]{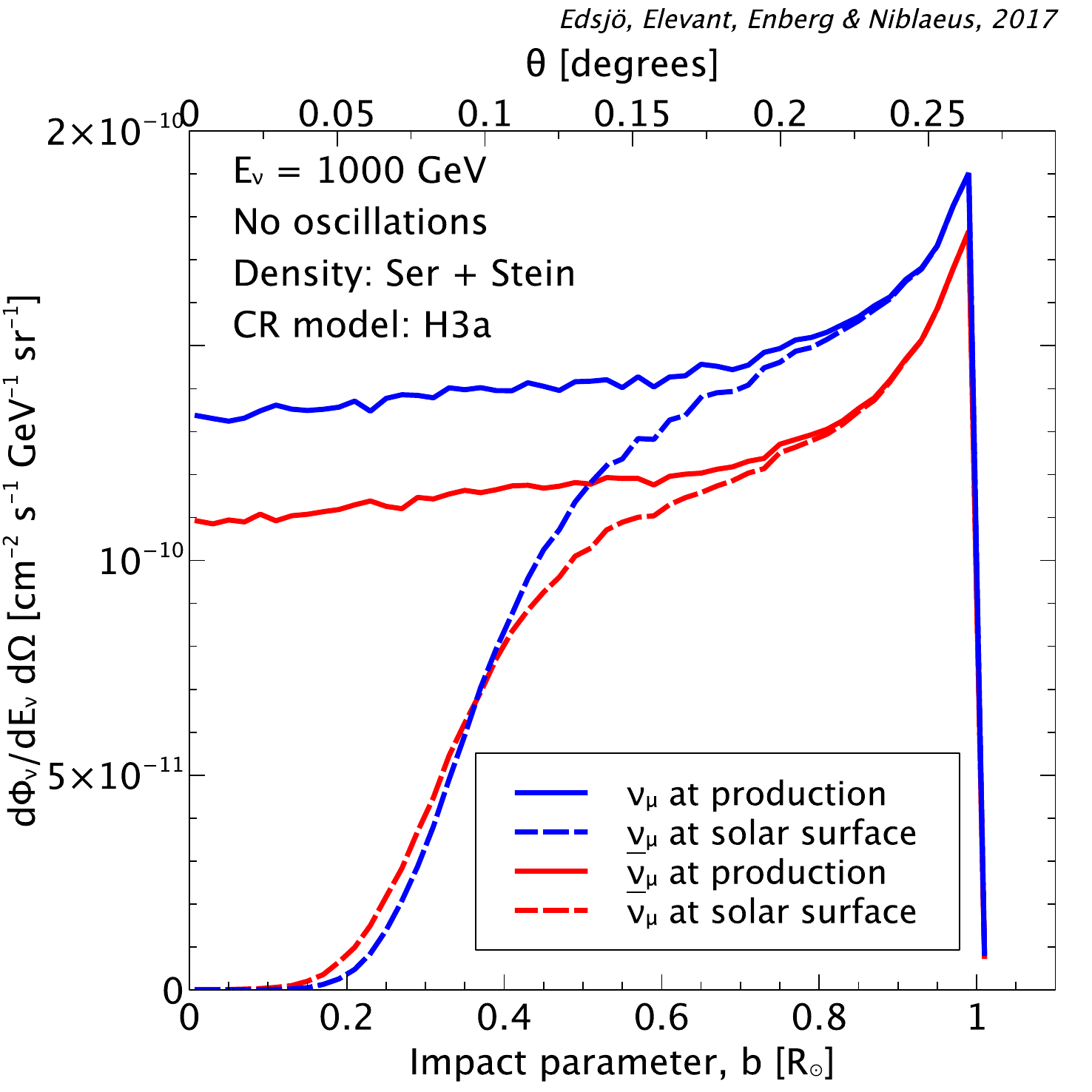}
\includegraphics[width=0.32\textwidth]{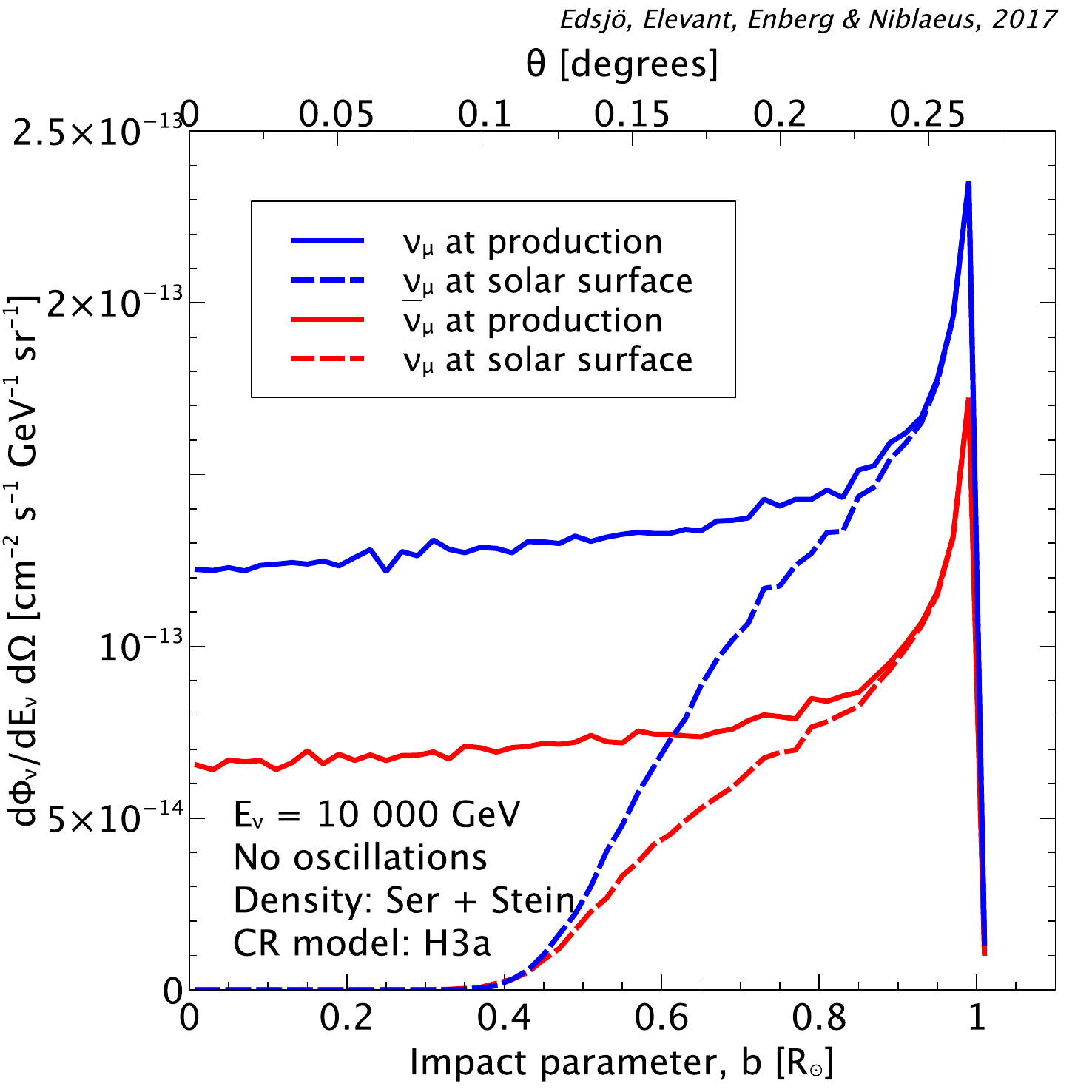}
\caption{The muon neutrino fluxes at production and at the surface of the Sun (after propagation through the Sun) for neutrino energies 100 GeV (left), 1000 GeV (middle) and 10\,000 GeV (right). In this figure, neutrino oscillations are not included to show the effect of attenuation only. Note that we here plot the fluxes and not $E_\nu^3$ times the fluxes.}
\label{fig:fluxes-vs-b}
\end{figure}

In figure~\ref{fig:fluxes-vs-b} we show the production fluxes and fluxes after passage through the Sun of muon neutrinos and antineutrinos for three different neutrino energies. For the sake of illustrating interaction effects we do not include oscillations in this figure. If we compare the neutrino fluxes at production (solid lines) to the ones after passage through the Sun (dashed lines), we see that we get a dip at low impact parameters. This is the effect of the attenuation that happens due to interactions when the neutrinos pass through the Sun, as we saw already in figure~\ref{fig:attenuation}.  As the density of the Sun is significantly higher in the centre, the effect is very strongly pronounced for low impact parameters.  We can also see that the effect of attenuation is higher for higher energies as expected.

We can also see how the production fluxes depend on the impact parameter. 
We see that for higher energies these are quite peaked at large impact parameters, which is expected as the density where the cascade happens is lower for these Sun grazing CRs, and hence the fluxes are higher. We also get a small contribution from muons decaying outside of the Sun at high impact parameters and high energies.\footnote{These are included in this figure even if they strictly speaking are produced between the Sun and the Earth.} The total flux from the Sun is obtained by integrating over the impact parameters including the fact that the solid angle is larger for large impact parameters. Hence, the high impact parameter part of these figures will be most important for the total flux from the Sun. So to summarise, we have three effects which all make the high impact parameters most important for our \sanu flux: i) interactions suppress low impact parameters, ii) CR interactions produce more neutrinos where the density is low, i.e.\ at high impact parameters and iii) the solid angle of the Sun is larger for large impact parameters.

At low $b$-values our fluxes to a large part agree with the previous calculation in IT96. At $b=1$ the fluxes in IT96 are significantly higher than at low $b$. We also see this effect, but not to the same extent. 

\subsection{Neutrino fluxes at a detector at Earth}

\begin{figure}
\centering
\includegraphics[width=0.40\textwidth]{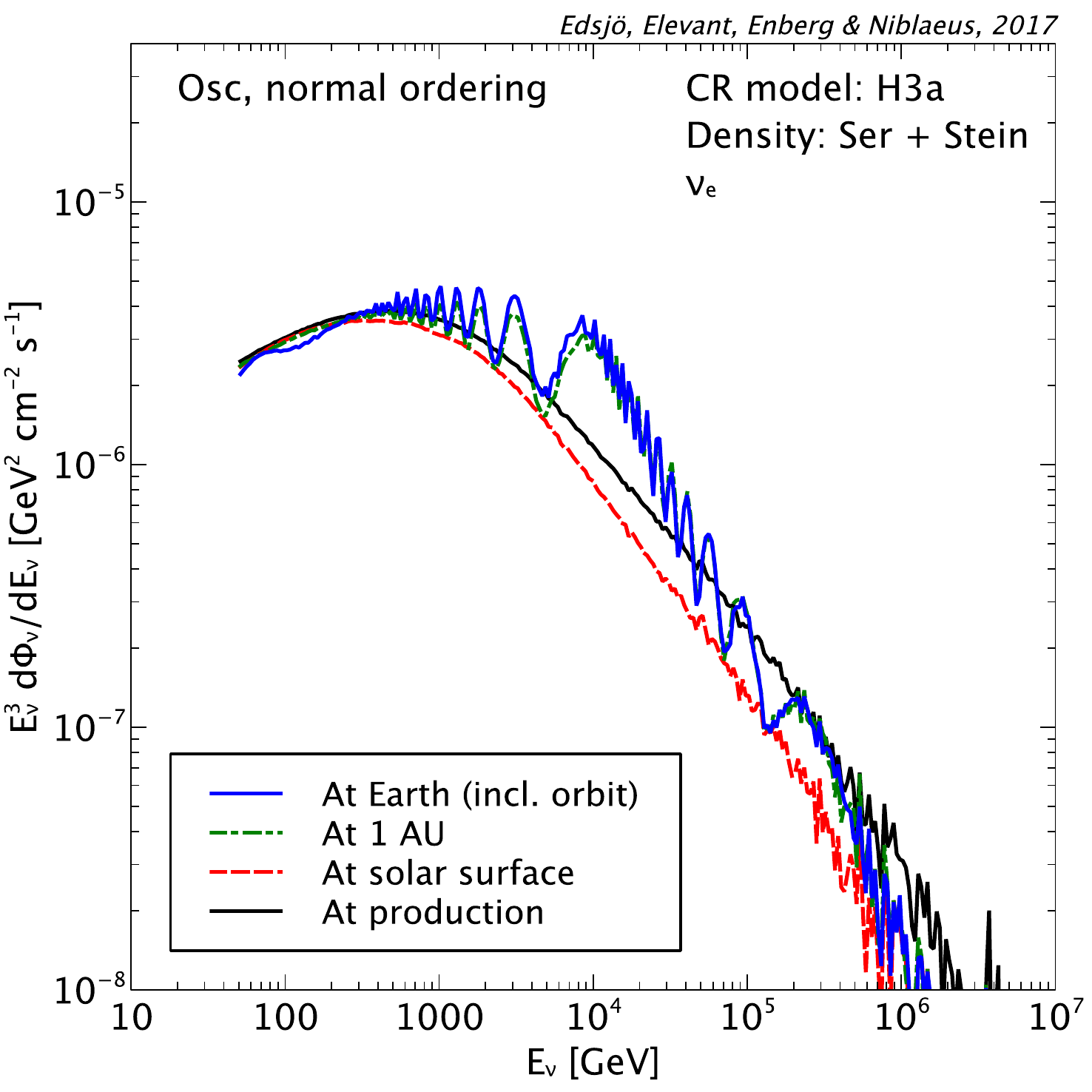}
\includegraphics[width=0.40\textwidth]{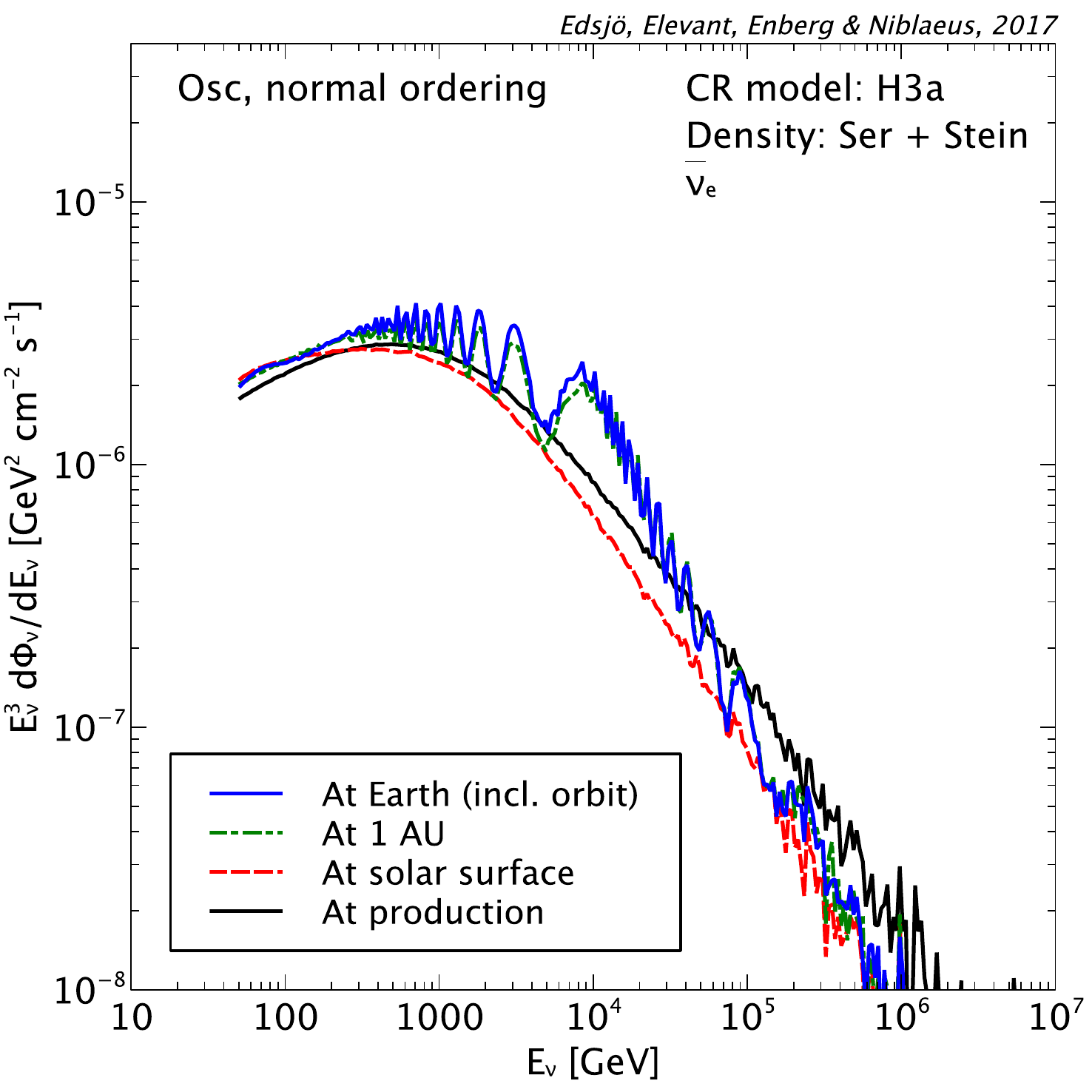}\\
\includegraphics[width=0.40\textwidth]{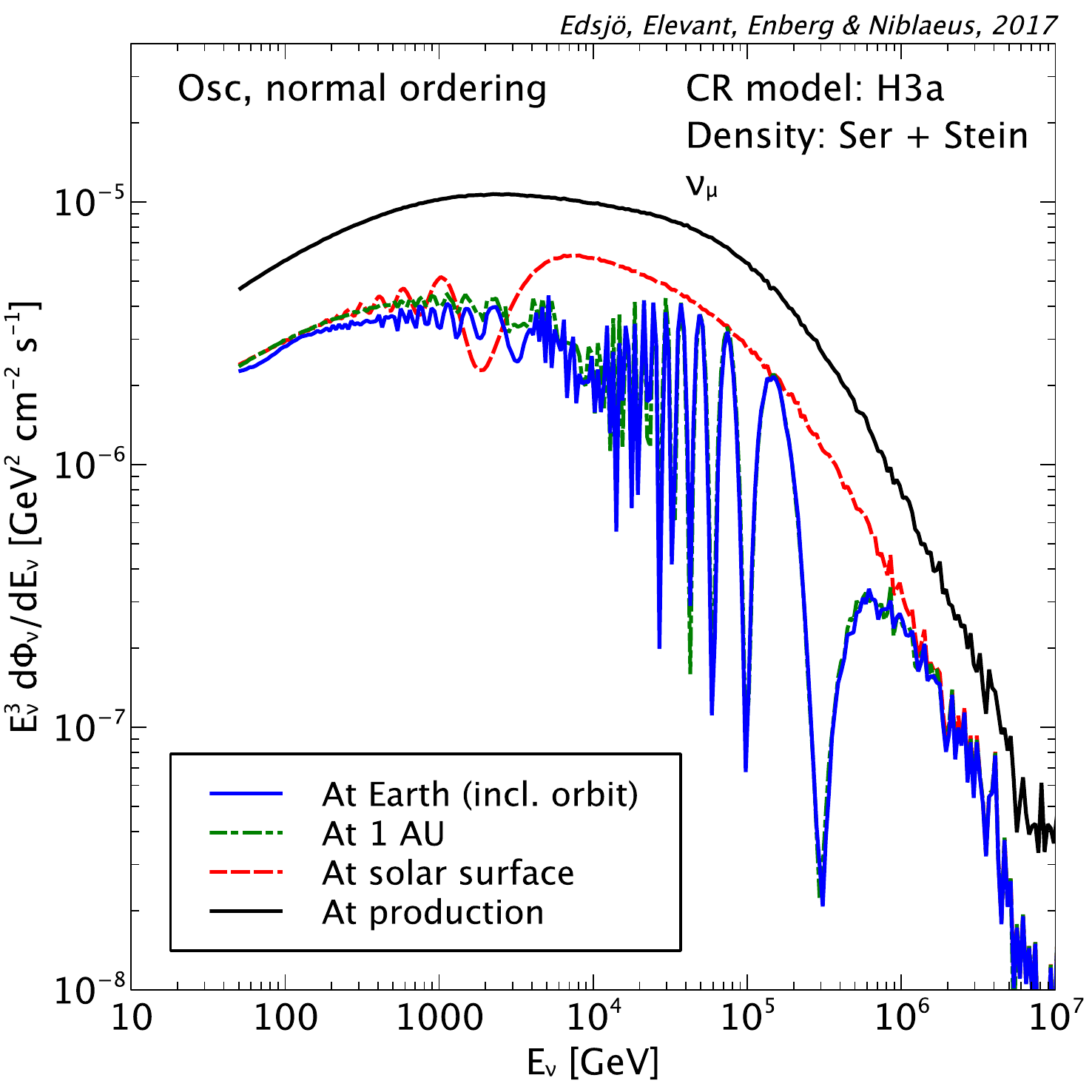}
\includegraphics[width=0.40\textwidth]{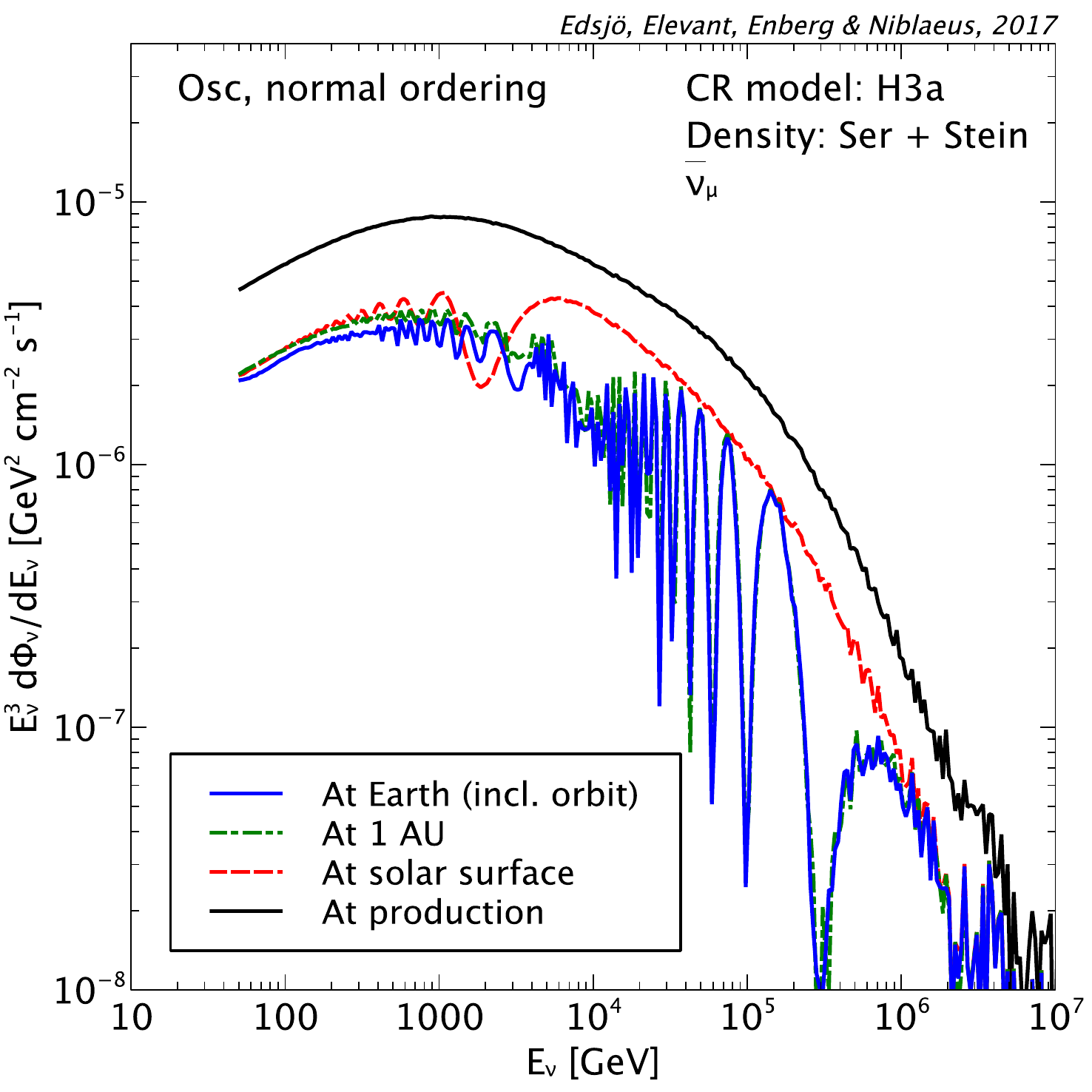}\\
\includegraphics[width=0.40\textwidth]{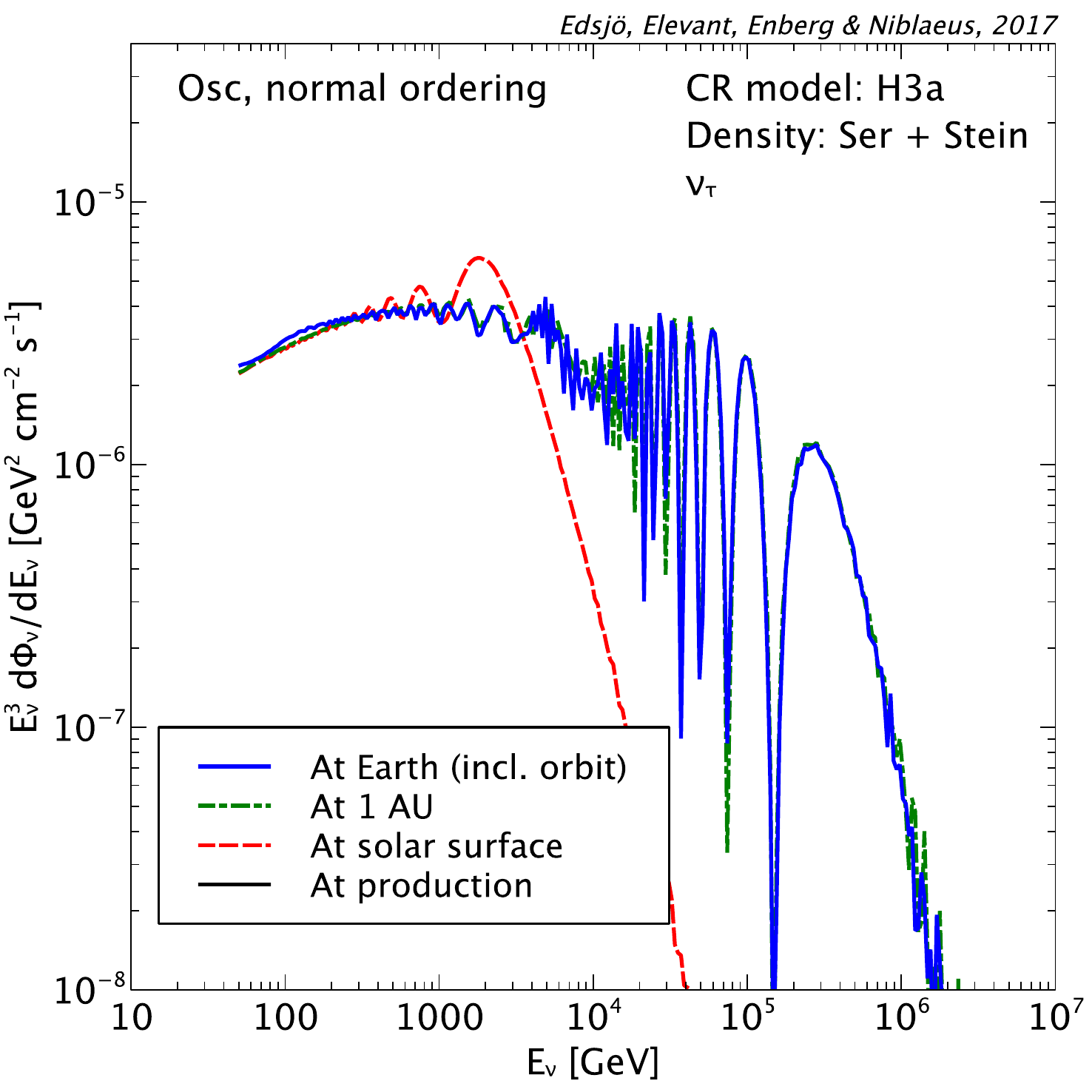}
\includegraphics[width=0.40\textwidth]{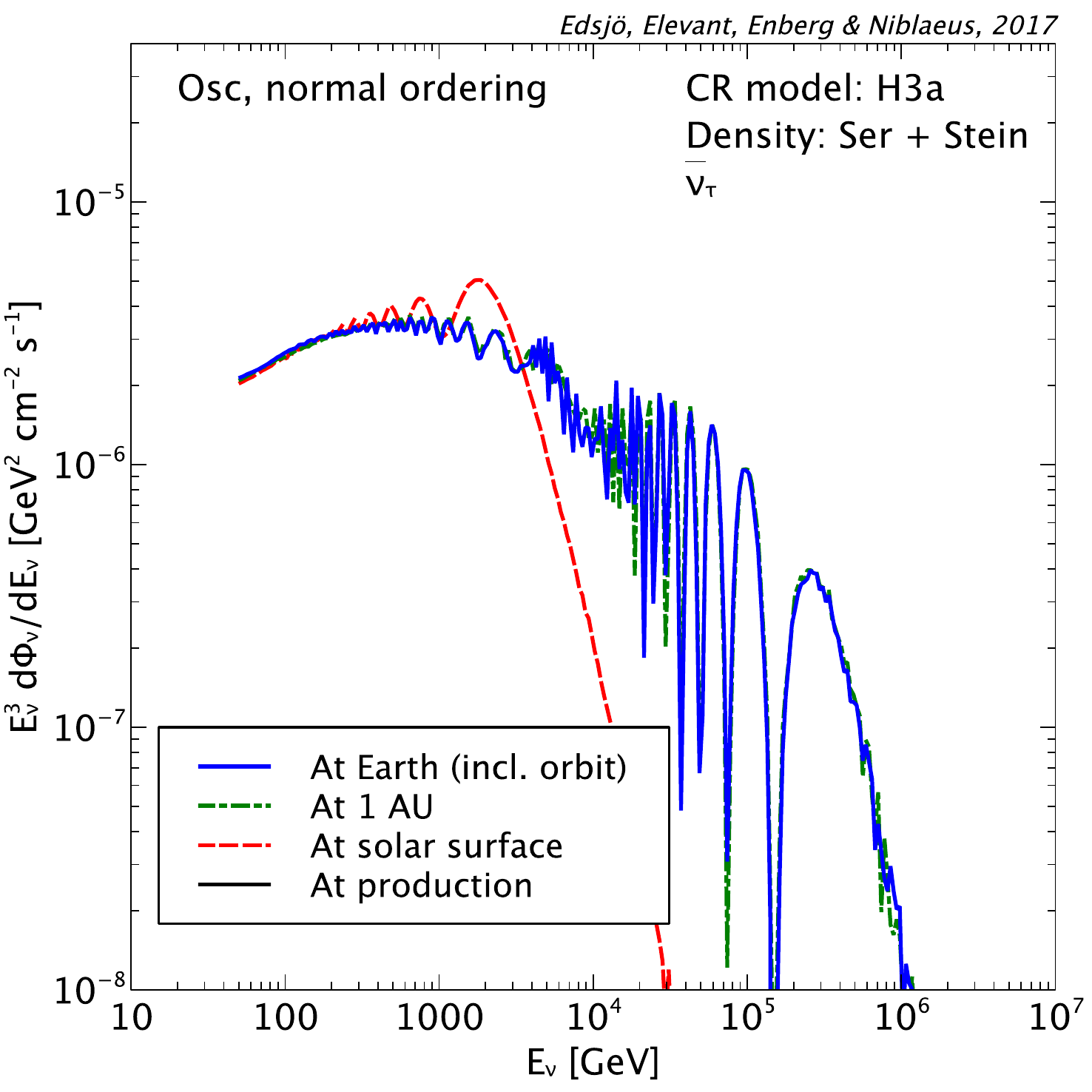}

\caption{The progression of neutrino fluxes from production, after passage through the Sun, at 1 AU from the Sun and finally at the detector (averaged over the austral winter). The plots show the $\nu_{e}$, $\nu_{\mu}$ and $\nu_\tau$ fluxes (integrated over the Sun) from top to down respectively. To the left we show neutrino fluxes and to the right anti-neutrino fluxes.}
\label{fig:progfluxes}
\end{figure}

\begin{figure}%
\centering
\includegraphics[width=0.32\textwidth]{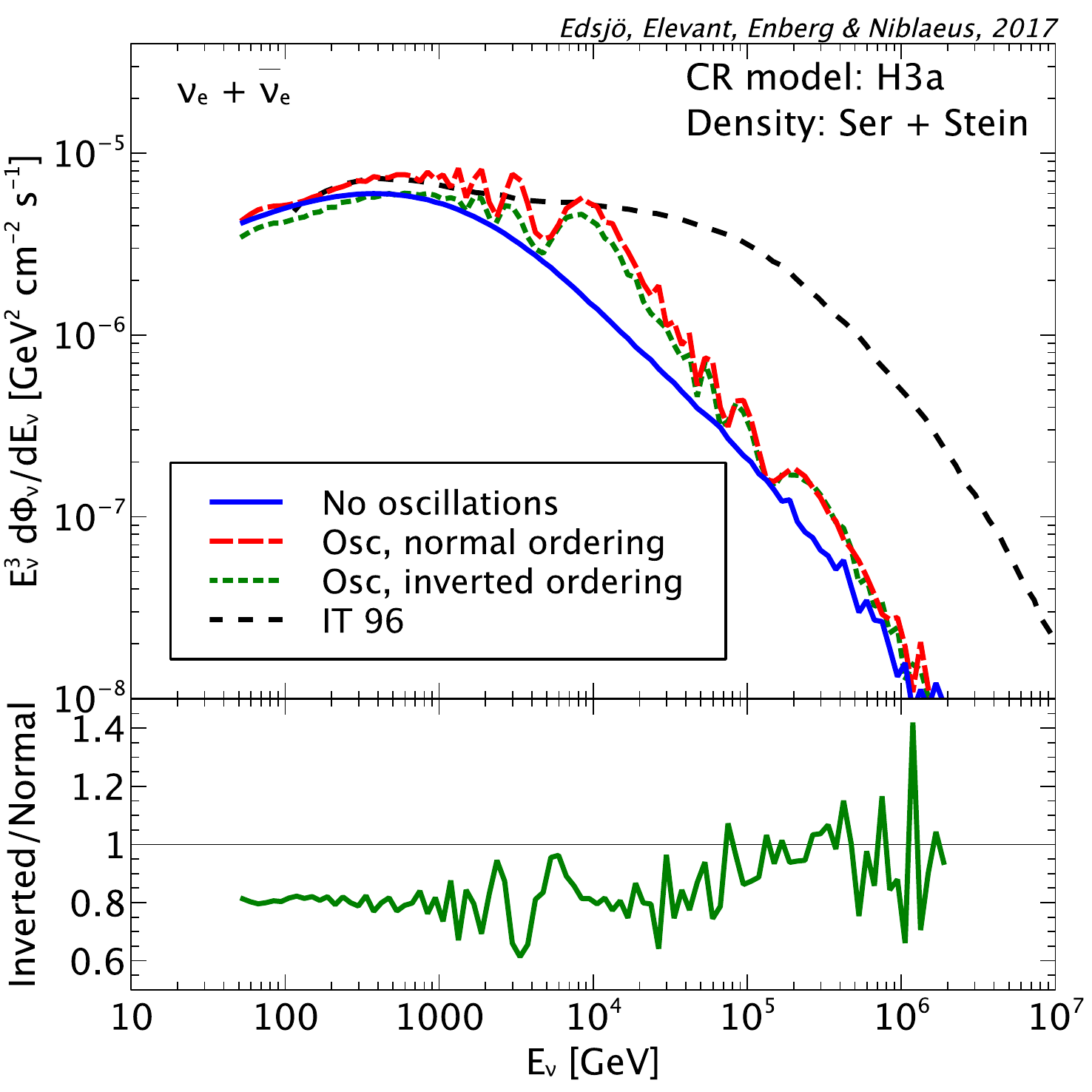}
\includegraphics[width=0.32\textwidth]{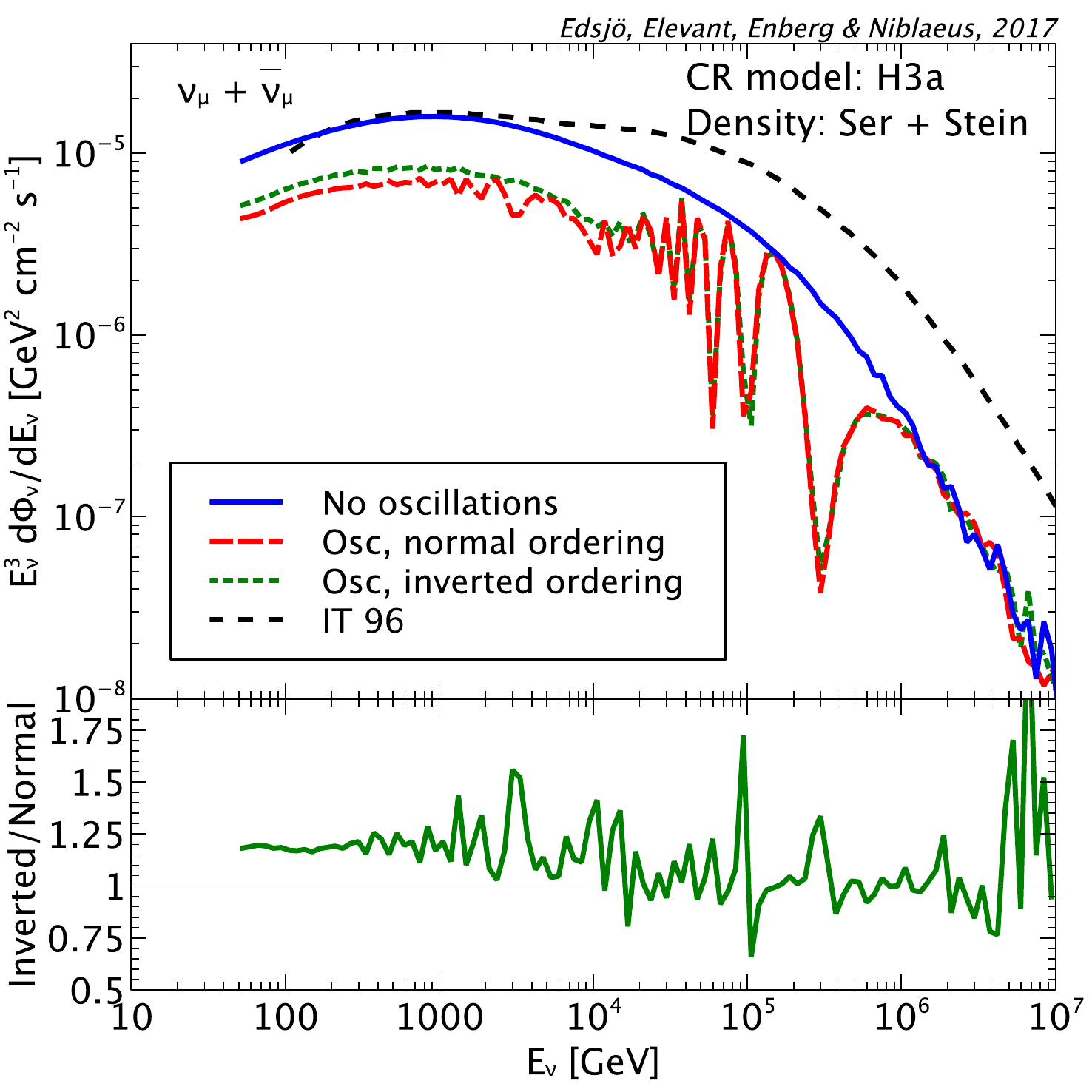}
\includegraphics[width=0.32\textwidth]{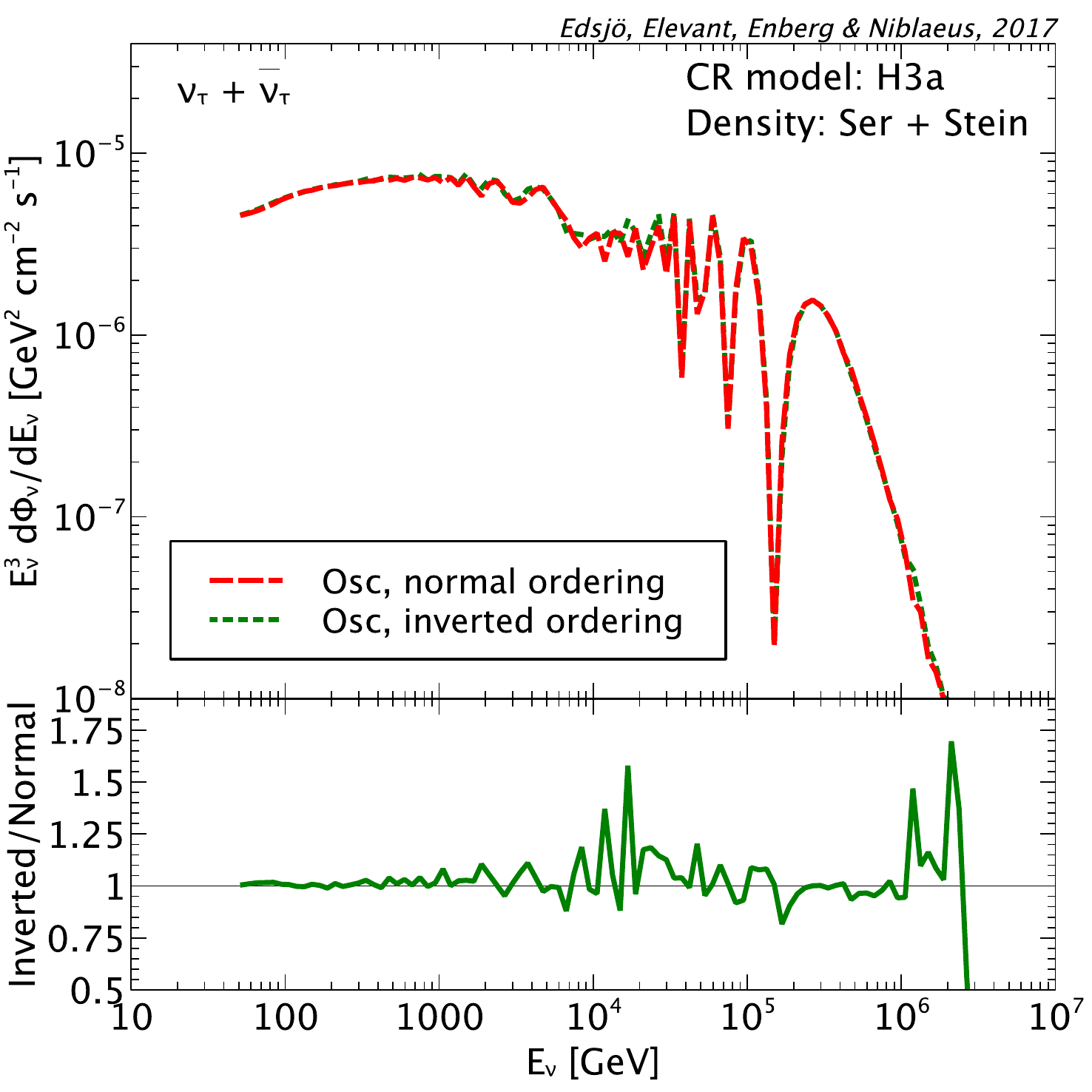}
\caption{The effect of different oscillation scenarios on the different neutrino flavour fluxes (at the detector, averaged over the austral winter). We show the sum of the electron neutrino fluxes (left), muon neutrino fluxes (middle) and tau neutrino fluxes (right). In the lower part of the plot the ratio between the inverted and normal ordering scenarios is shown. Note that for tau neutrinos, the fluxes without oscillations are below the lowest boundary of the plot and are hence not shown. For electron and muon neutrinos we also show the fluxes from IT96 \protect\cite{Ingelman:1996mj}}
\label{fig:nuoscfluxes}
\end{figure}

In figure~\ref{fig:progfluxes} we show the neutrino fluxes at production, after propagation through the Sun, at 1 AU from the Sun and finally propagated to the detector at Earth. The difference between the results at 1 AU and at the detector is that the detector fluxes are averaged over the austral winter (from vernal to autumn equinox). This averaging will due to the eccentricity of the Earth's orbit wash out some of the oscillation effects at lower energies. 

After propagating through the Sun we can see from the red curves in figure~\ref{fig:progfluxes} that some neutrinos have been lost due to interactions, especially at higher energies. We also see that on these length scales oscillations essentially only affect muon and tau neutrinos. The oscillation length $\lambda_{21}$ due to $\Delta m_{21}^2$ is in accordance with eq.~\eqref{eq:osc-length1} large compared to $R_\odot$ and hence ineffective at these energy and length scales. Therefore oscillations from muon neutrino into electron neutrinos are insignificant. Muon neutrinos oscillations into tau neutrinos, coming from the oscillation lengths $\lambda_{31}$, $\lambda_{32}$, are significant at energies below $\sim \SI{e3}{\giga\electronvolt}$. Below  about  $\sim \SI{e2}{\giga\electronvolt}$ the neutrinos are incoherent since from eq.~\eqref{eq:osc-length2} we have $\lambda_{31},\lambda_{32}\gg R_\odot$ and oscillations average out. At very high energies, also  $\lambda_{31}$ and $\lambda_{32}$ are long compared to  $R_\odot$ and oscillations do not develop, leading to a tau neutrino flux that is very small and does not differ from the very small production flux.

In the fluxes after passage through the Sun  we can see some effects of matter oscillations. There is a slight enhancement in the  flux of $\bar{\nu}_e$ at around \SI{100}{\giga\electronvolt} that is due to matter effects, but overall the oscillation effects on the fluxes are well approximated by vacuum oscillations.

At the Earth the distance travelled is long compared to all oscillation lengths for energies below  a few hundred \si{\giga \electronvolt} and the  oscillations average out, resulting in a ratio of about equal fluxes for all three flavours. At high  energies we can now see the effect of oscillations due to $\lambda_{21}$ in the electron neutrino flux and the effect of all oscillation lengths for the muon and tau neutrinos. For the latter two, $\lambda_{21}$ oscillations now appear in the region around $\sim \SI{e3}{\giga\electronvolt}$ and the $\lambda_{31} \approx\lambda_{32}$ oscillations at higher energies around  $\sim \SI{e4}{\giga\electronvolt}$ with muon oscillation into tau neutrinos now efficient up to the highest energies. Thus the tau neutrino flux at the Earth is almost entirely due to oscillations.

In figure~\ref{fig:nuoscfluxes} we show the effect of different oscillation parameters. The `No oscillations' model is of course not physical, but just shown for comparison. The normal and inverted ordering refers to the best fit neutrino oscillation parameters for these two cases. The main effect of inverted ordering compared to normal ordering is to increase the muon neutrino fluxes below $10^4$ GeV and correspondingly decrease the electron neutrino fluxes. The tau neutrino fluxes are not affected by the mass ordering. We attribute the difference from mass ordering to the different values of the best fit parameters. For electron and muon neutrinos we also compare with the IT96 \cite{Ingelman:1996mj} results. We note that without oscillations our results agree fairly well with IT96 at low energies, whereas at higher energies our results are lower. It is hard to know exactly what causes this difference, as it can come from many different sources (cosmic ray model, atmospheric interaction model, solar model, etc). We have though compared with their production fluxes for different impact parameters (their figure~1). Our production fluxes are in reasonable agreement, except at $b=1$ where they get significantly more neutrinos. We also get more neutrinos for large impact parameters, but not to the extent IT96 gets them. We also note that IT96 calculates the fluxes at three impact parameters ($b=0$, $b=2/3$ and $b=1$) and then interpolate between these to get the total flux from the Sun. As the flux is so much higher at $b=1$, the way the interpolation is done will largely affect the result. We have generated \mceq tables for more values of $b$ (especially close to 1) to make sure we get small interpolation errors and then draw events for all $b$. Our integration over the Sun should therefore be more accurate.

\begin{figure}%
\centering
\includegraphics[width=0.32\textwidth]{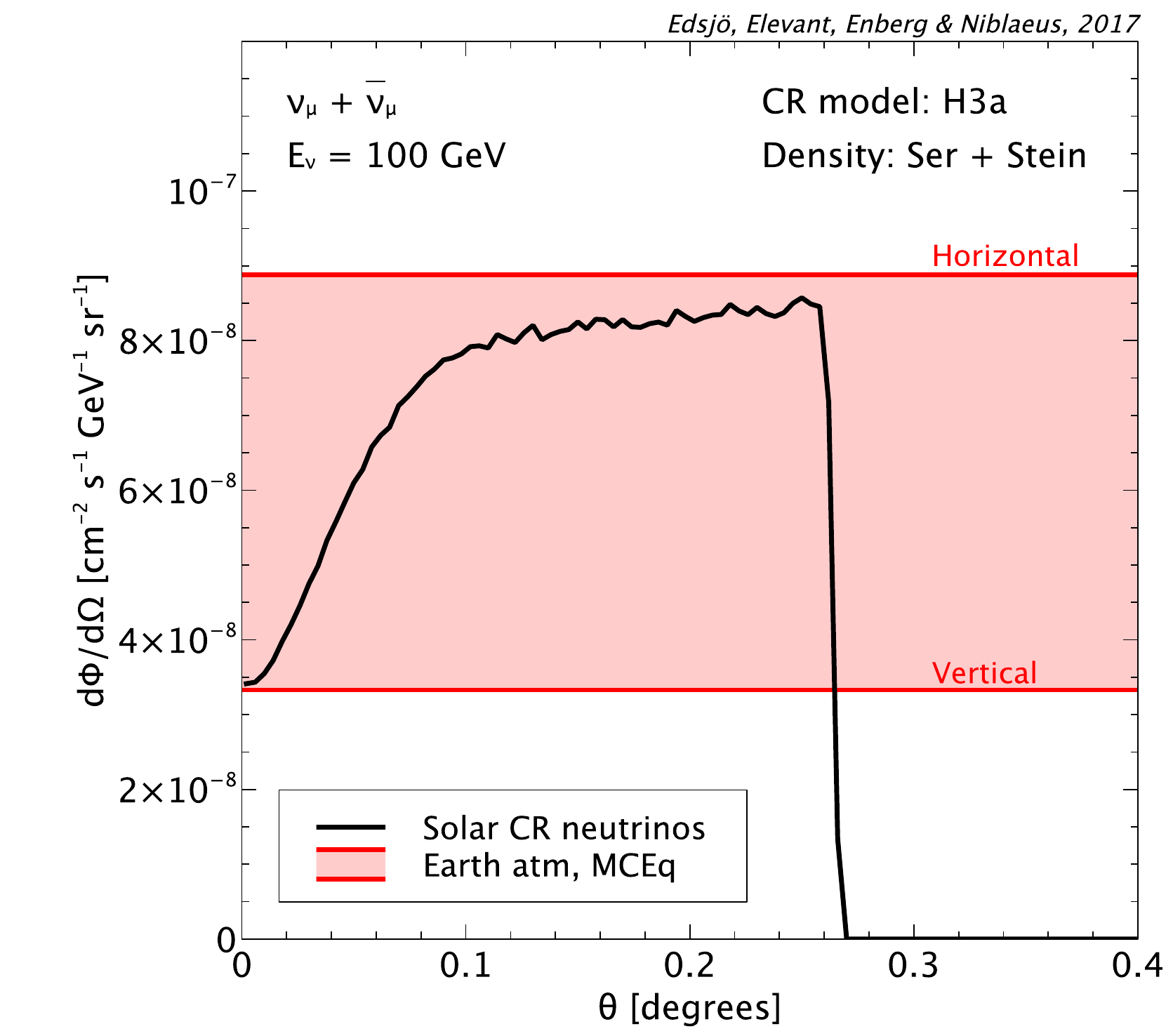}
\includegraphics[width=0.32\textwidth]{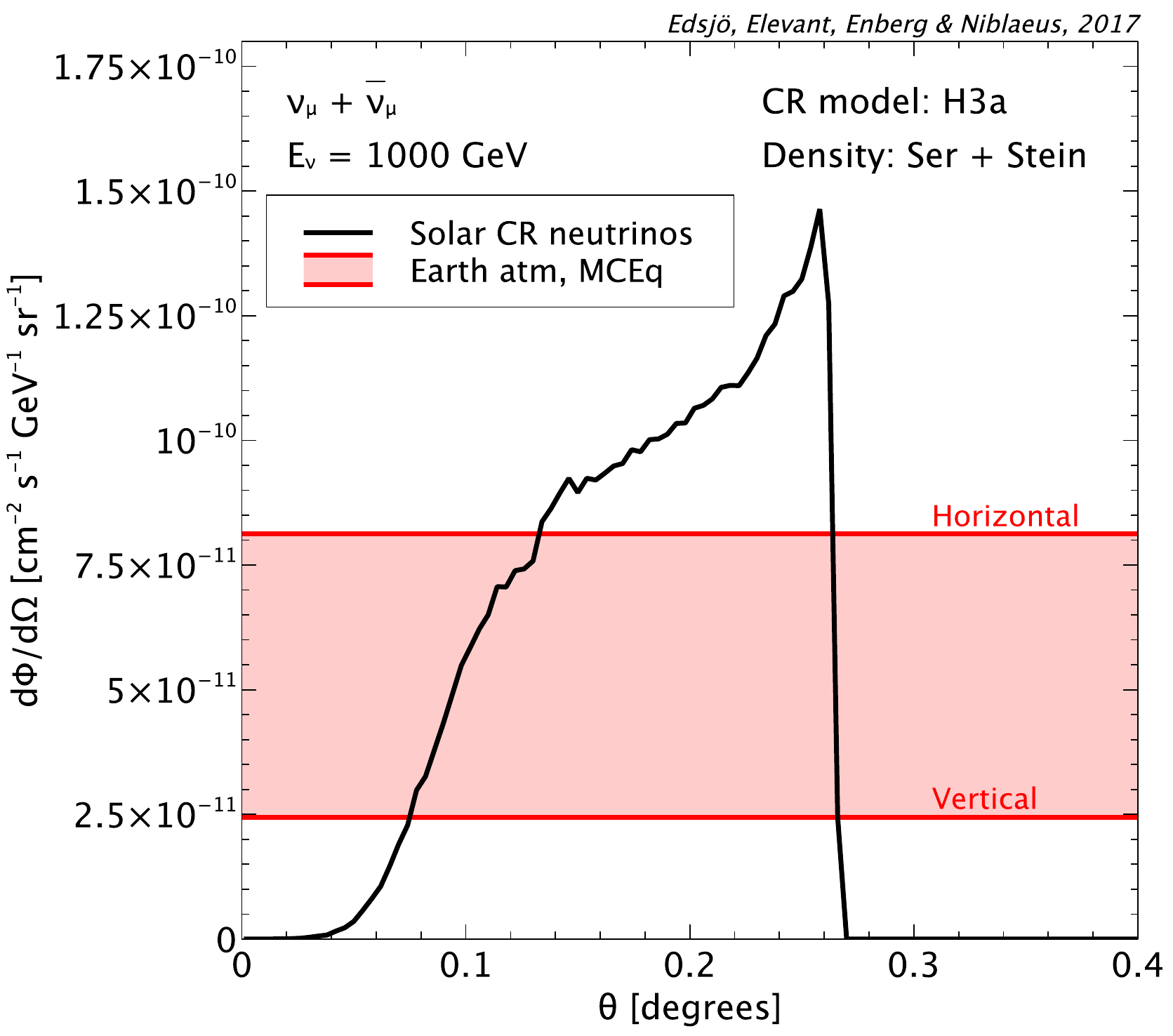}
\includegraphics[width=0.32\textwidth]{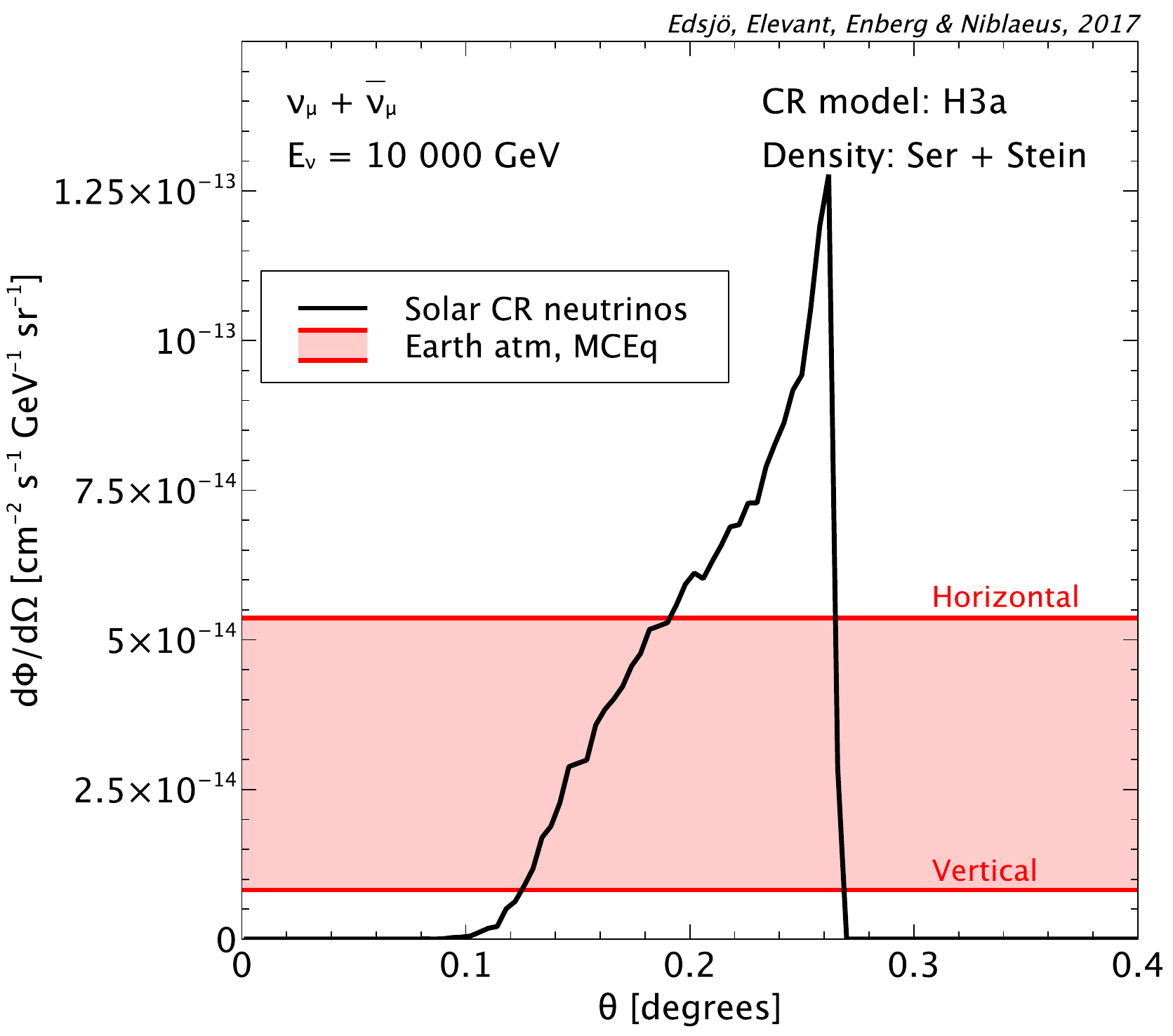}
\caption{The muon neutrino flux from the Sun compared to the Earth atmospheric neutrino background (both are given per solid angle). The Earth atmospheric fluxes are given both for horizontal and vertical fluxes (with a shaded red region in between). For an actual detector, the Earth atmospheric fluxes will be between these two extremes (and slightly reduced in the direction of the Sun due to the Sun blocking some cosmic rays). The Sun and Earth fluxes are both calculated with \mceq with the same set of parameters.}
\label{fig:nuatmfluxes}
\end{figure}

In the previous figures we showed the fluxes integrated over the Sun, but it is also interesting to look at the angular distributions and compare to the Earth atmospheric neutrino fluxes. One way to view this is that the cosmic rays hit an atmosphere and produce cascades and eventually neutrinos. If it were not for interactions, oscillations and atmospheric differences we would expect to get essentially the same flux of neutrinos (per solid angle) from the Sun's and Earth's atmospheres. In principle, the Sun blocks some cosmic rays to reach the Earth and we would naively expect to get a reduction of Earth atmospheric neutrinos in the direction of the Sun, and an equal increase from the cosmic ray interactions in the Sun. However, including atmospheric differences (the Sun's atmosphere is considerably less dense), interactions and oscillations, this no longer holds true and the solar cosmic ray neutrinos could be both larger or smaller than the Earth atmospheric ones.
 
In figure~\ref{fig:nuatmfluxes} we show the differential muon neutrino ($\nu_\mu$ + $\bar{\nu}_\mu$) fluxes (per solid angle) from the solar cosmic ray neutrinos (i.e.\ our calculation in this paper) and compare with the Earth atmospheric neutrinos. Our shown fluxes in this plot are very similar to those shown in figure~\ref{fig:fluxes-vs-b}, with the difference being that here we show the fluxes propagated all the way to the detector and we include oscillations. An actual detector will of course be at a given latitude and the Sun will then be in a range of directions on the sky, so depending on the detector location and time of day and year the actual Earth atmospheric background will be somewhere between the lower (vertical) and upper (horizontal) limit. We have in this figure not included the solar cosmic ray blocking effect on the Earth's atmospheric fluxes as it is complicated to model given the effect of magnetic fields in the Solar System (naively one would expect the Earth atmospheric neutrino fluxes to drop to zero below 0.26$^\circ$). We can see that the solar cosmic ray neutrinos are of the same order or larger than the Earth atmospheric background, but we get a dip in the centre due to the attenuations through the Sun. If we compare the different energies in Figure \ref{fig:nuatmfluxes} we also see the the \sanu fluxes are (relatively speaking) higher than the Earth's atmospheric neutrino fluxes especially for higher energies. This can be understood from the density of the solar atmosphere being much lower than the Earth's atmosphere. This means that especially at high energies, the unstable cascade particles have time to decay before they interact, whereas in the Earth they are much more likely to interact.

\subsection{Neutrino-induced muon fluxes at a detector at Earth}

\begin{figure}%
\centering
\includegraphics[height=0.45\textwidth]{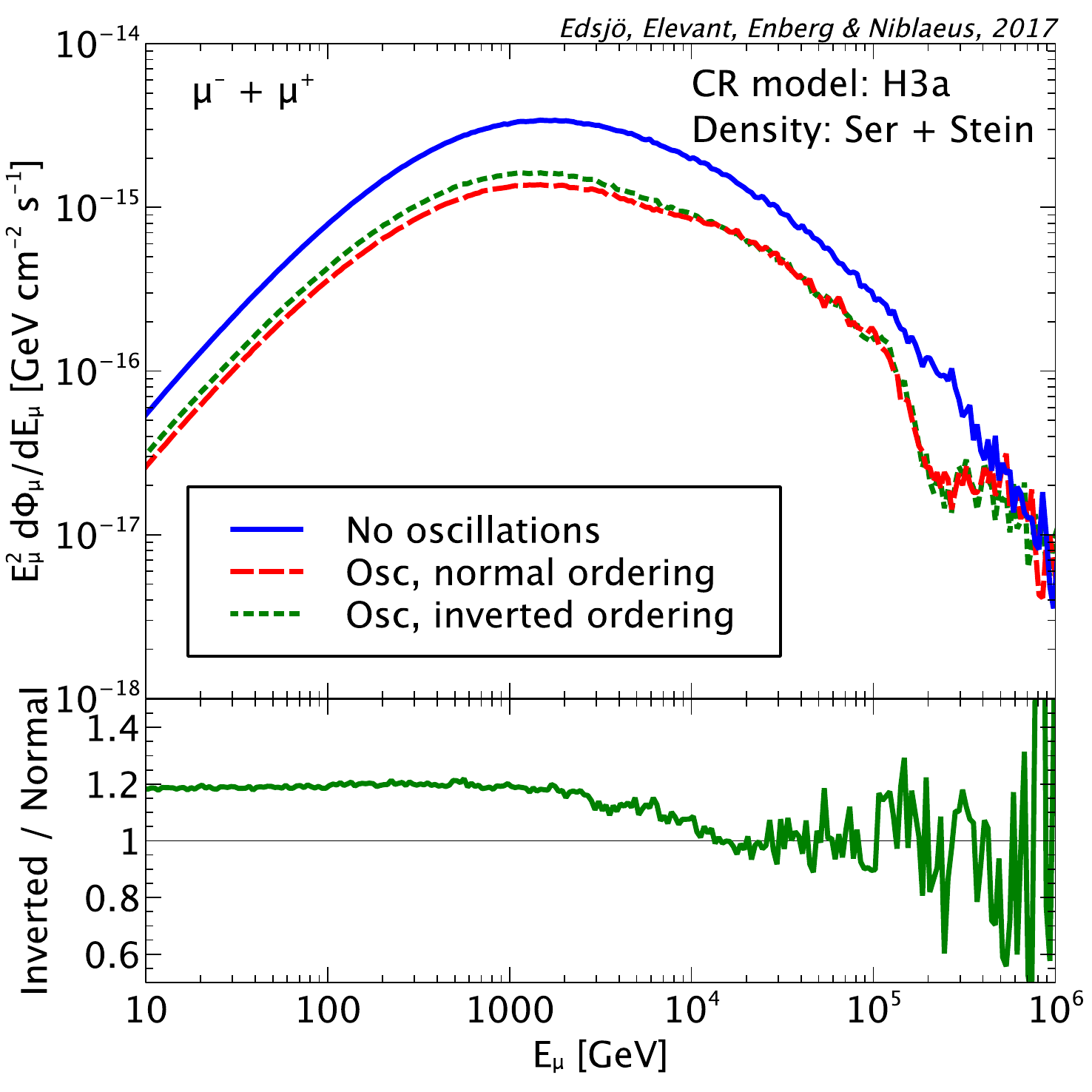}
\includegraphics[height=0.45\textwidth]{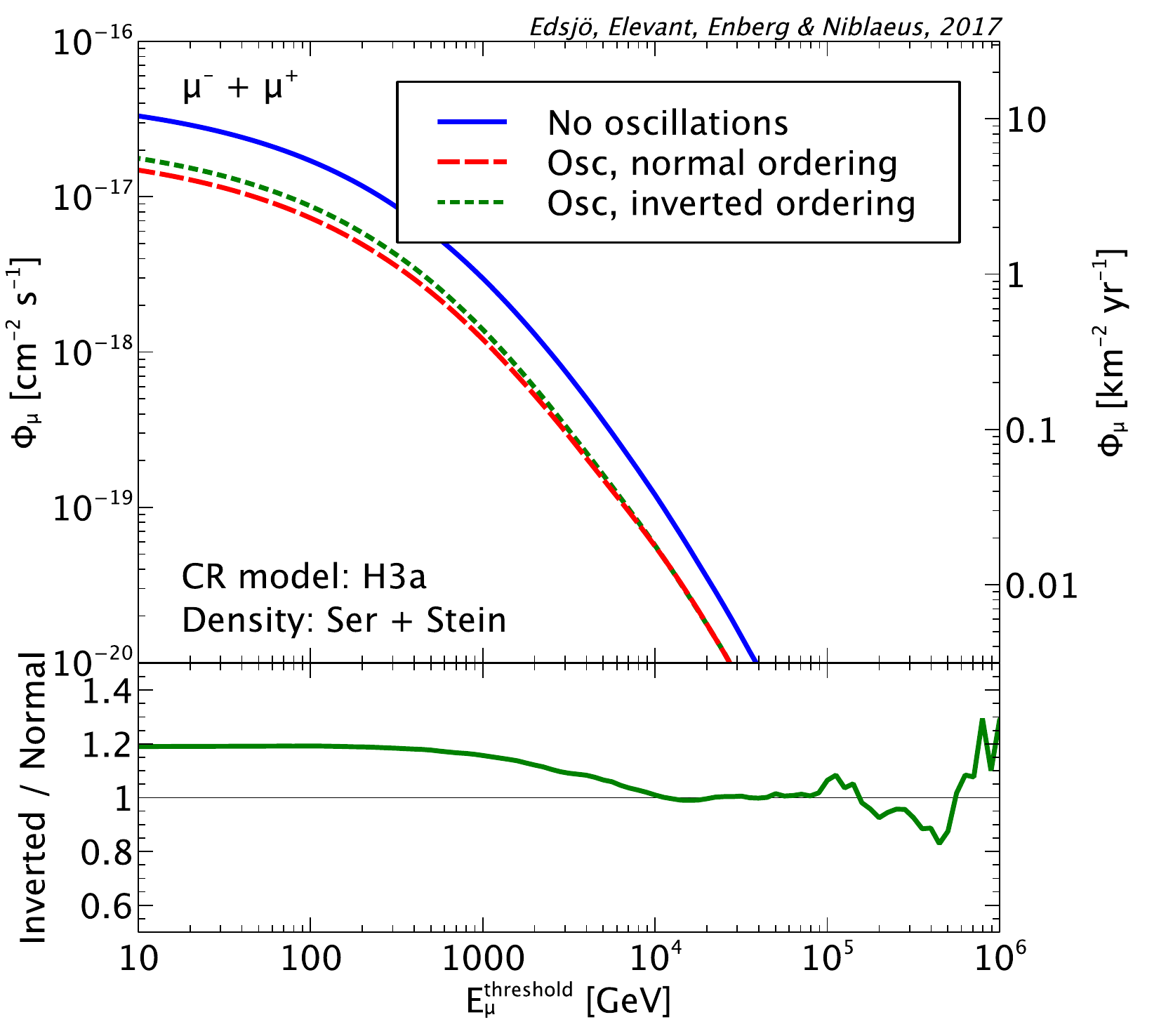}

\caption{The flux of neutrino-induced muons at the detector (averaged over the austral winter). At the bottom, the ratio between the inverted and normal oscillation scenarios is shown. To the left we show differential fluxes (note that these fluxes are shown as $E_\mu^2 d\Phi_\mu/dE_\mu$) and to the right integrated fluxes above an energy threshold.}
\label{fig:muoscfluxes}
\end{figure}

Instead of looking at the neutrinos directly we can let these interact and look at the muons that are produced. The muon flux will of course be lower as most neutrinos (at least at lower energies) do not interact. With the \wimpevent part of \wimpsim we have calculated the neutrino-induced muon flux at a detector in ice at the South Pole. In figure~\ref{fig:muoscfluxes} we show the resulting muon ($\mu^- - \mu^+$) fluxes at the detector (to the left differential in energy and to the right integrated over an energy threshold). To get these fluxes we have let the muons lose energy after the neutrino-nucleon interaction, i.e.\ we show the muon flux at a plane perpendicular to the Sun at the detector. We see that the effect we saw earlier, that inverted neutrino mass ordering gives higher fluxes, still remains. In the right figure with the integrated fluxes, one can also get a rough estimate of the event rates in a neutrino telescope. A detector with an effective area of 1 km$^2$ and a muon energy threshold of 100 GeV would e.g.\ see a flux of 2.3 muons ($\mu^- + \mu^+$) per square kilometre per year. In the next section, we will calculate event rates more accurately using the effective areas for different energies instead.

\begin{figure}%
\centering
\includegraphics[width=0.49\textwidth]{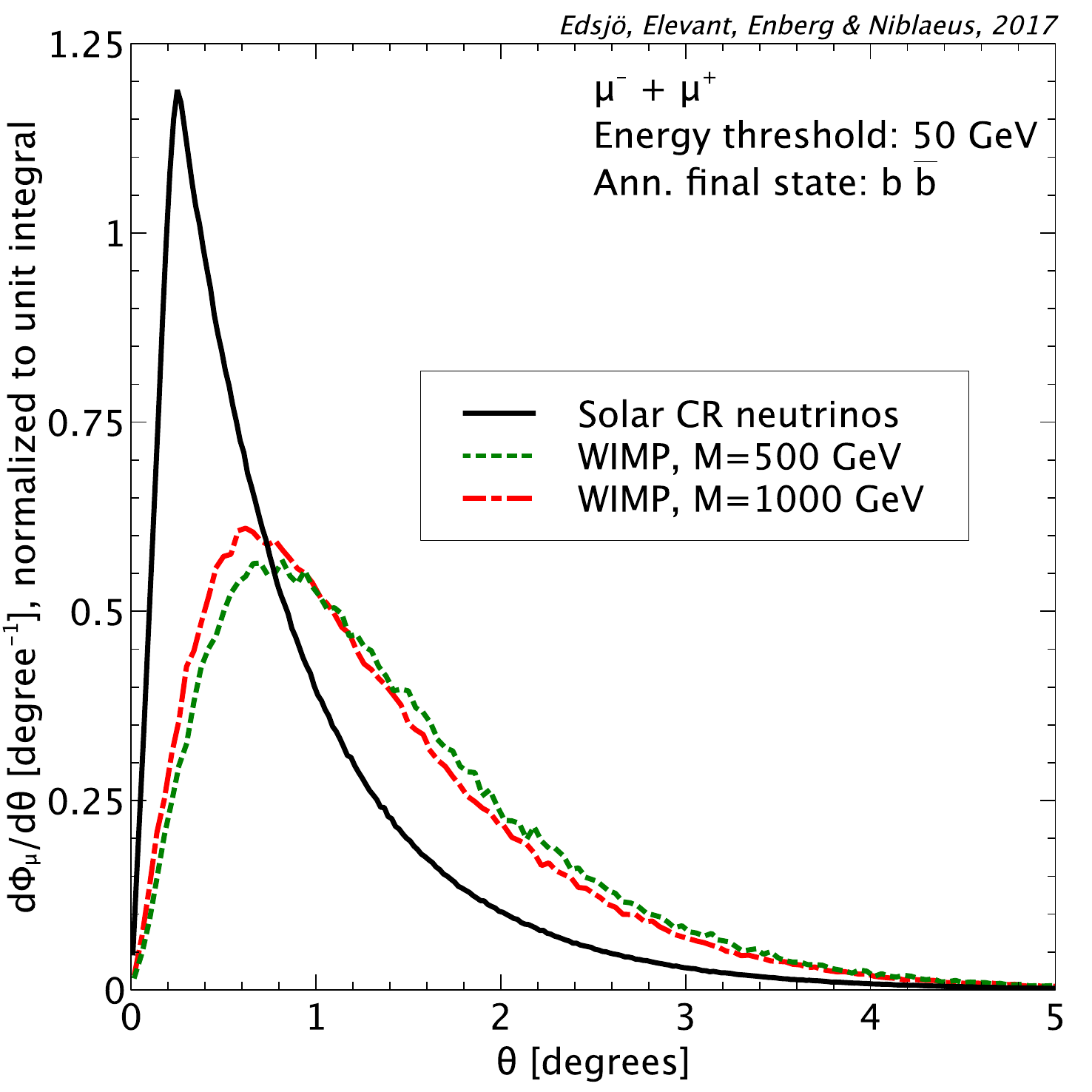}
\includegraphics[width=0.49\textwidth]{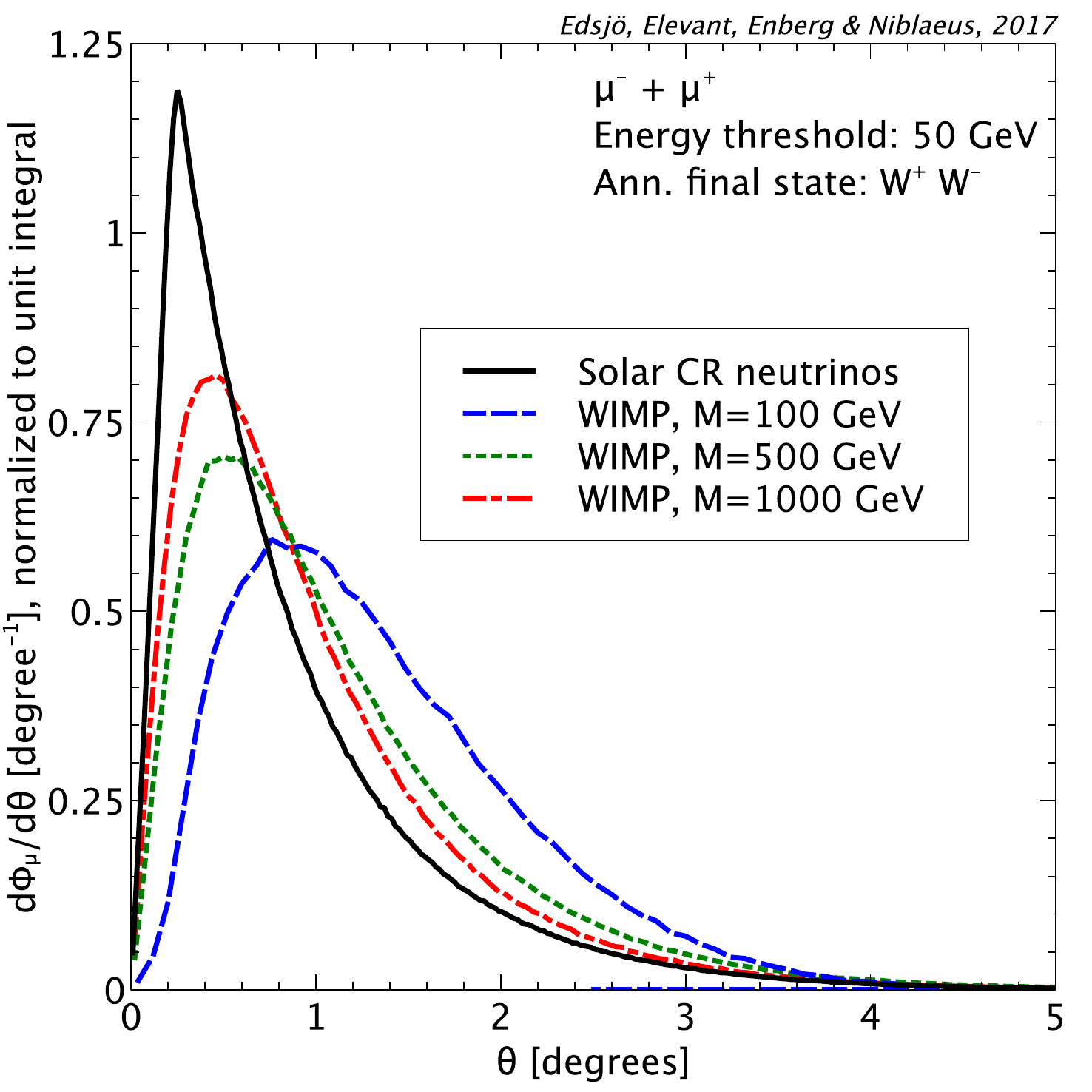}
\caption{The angular distribution of the neutrino-induced muon flux at the detector with a threshold of 50 GeV on the muon energies. For comparison we show the angular distributions of a WIMP annihilation signal of different masses for annihilation into $b\bar{b}$ (left) and $W^+ W^-$ (right). In this figure, we have integrated over the azimuthal angle so that the fluxes are differential in the angle with respect to the Sun, $\theta$.}
\label{fig:angfluxes}
\end{figure}

\begin{figure}%
\centering
\includegraphics[width=10cm]{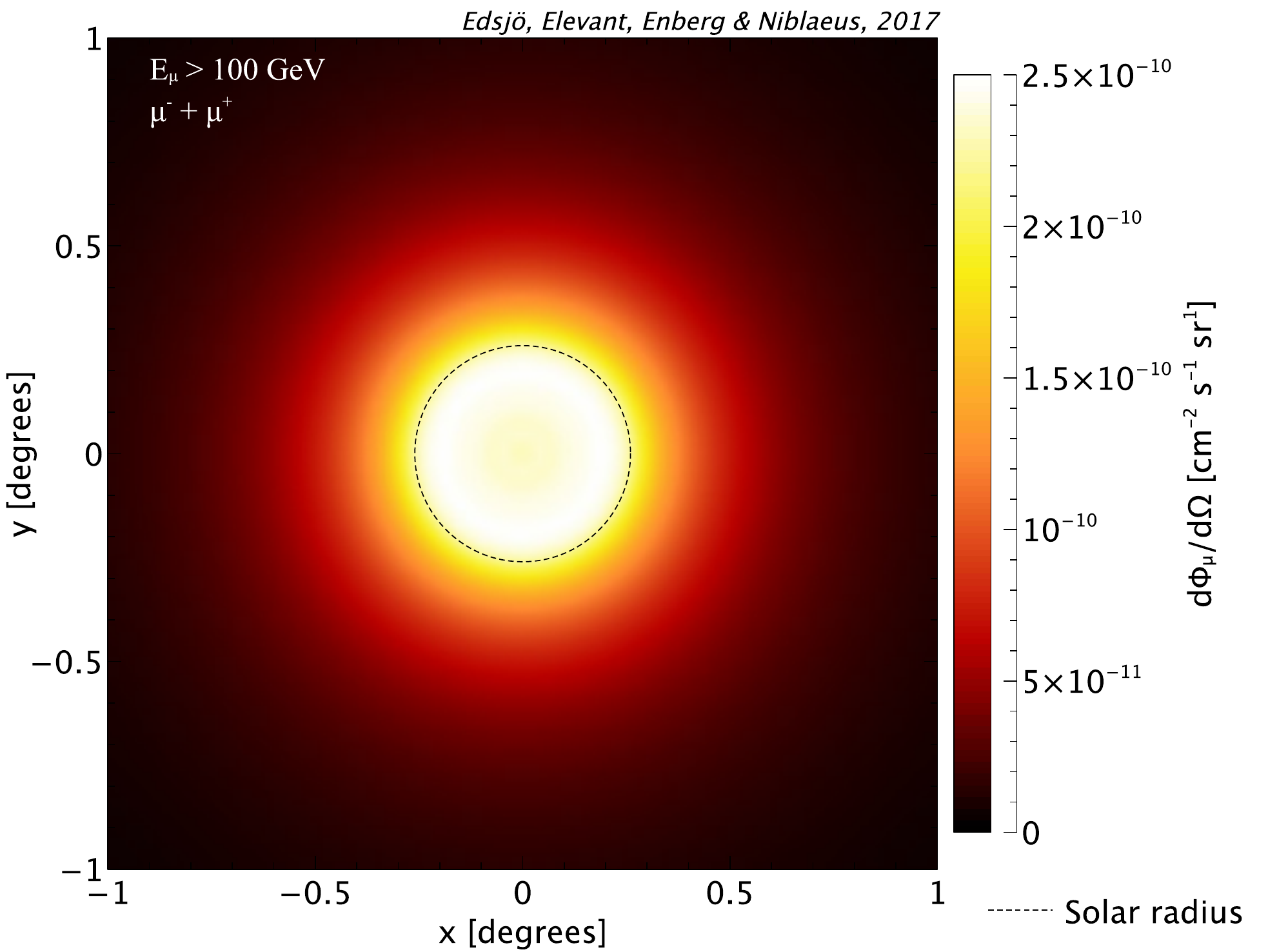}
\caption{The Sun as it could be seen in neutrino-induced muons at a neutrino telescope. Note the dip in the centre.}
\label{fig:sunimage}
\end{figure}

In figure~\ref{fig:nuatmfluxes} we looked at the angular distributions of neutrinos, but as these angles are not directly observable, we now instead focus on the angular distributions of the muons. In figure~\ref{fig:angfluxes} we show the fluxes of neutrino-induced muons at the detector integrated above a muon energy of 50 GeV. Compared to the neutrino angles (that were at most 0.26$^\circ$, half the angular diameter of the Sun), these angles now include also the neutrino-nucleon scattering angle and the deflection of the muons due to multiple Coulomb scattering. Hence, these angles are slightly bigger.

We have mentioned earlier that these solar cosmic ray neutrinos will be an (essentially) irreducible background for searches for neutrinos from dark matter annihilations in the Sun. Both the solar cosmic ray neutrinos and the neutrinos from dark matter annihilations come from the Sun, but as the energy distribution is different, so will the angular distribution be (mostly from the fact that the neutrino-muon angle at the deep inelastic scattering goes like $1/\sqrt{E_\nu}$). In figure~\ref{fig:angfluxes} we also show the angular distributions for some dark matter models, so called WIMPs. These are also calculated with \wimpsim where Pythia \cite{Sjostrand:2006za} is used to calculate the annihilation spectrum of neutrinos from WIMP annihilations. We show these fluxes as differential in the angle from the Sun, $\theta$, as this is close to what one would experimentally cut on if trying to separate these two distributions. In principle, we could also cut on energy, but the energy estimate of muons at these energies is very poor. We have not included the experimental error on the angle in this figure, if we would it would smear the distributions further. Even if the angular distributions are different we argue that they are quite close which will make discrimination difficult, hence the solar cosmic ray neutrinos will be a background for dark matter searches from Sun.

Finally, for illustration we show in figure~\ref{fig:sunimage} how the Sun wold look like if we could see the neutrino-induced muons from solar cosmic ray interactions. Even if the neutrino-nucleon scattering angle smears the neutrino fluxes somewhat, we still see the ring like signature of this signal with a slight dip in the centre.

\subsection{Events in neutrino telescopes}
\label{sec:events}

To estimate how many events we could expect in a neutrino telescope, we focus on IceCube and use the estimated effective areas for searches looking for a neutrino excess from the Sun. We have used two recent estimates of the neutrino effective area, one from their IC-79 study \cite{Aartsen:2016exj} and one from their recent 3-year study \cite{Aartsen:2016zhm}, we will call these \emph{IC-79} and \emph{IC3} below. These two estimates use slightly different event selection criteria and present their effective areas in different ways (separate for $\nu_\mu$ and $\bar{\nu}_\mu$ in the first one and combined in the second). For \emph{IC-79} we use the highest effective area of the three selection criteria SL, WL and WH for each energy. For \emph{IC3} we use the DeepCore selection for low energies and the IceCube selection for higher energies. The events per year (of lifetime of the detector) are given in Table~\ref{tab:icevents}. The event rates for different cosmic ray and density models differ only by about 1--1.5\% from the values listed in the table.

\begin{table}[h!]
\caption{Number of events per year (lifetime) in IceCube. The neutrino flux above 50 GeV has been used in calculating these event rates for \emph{IC-79} and \emph{IC3}.}
\begin{center}
\begin{tabular}{lrr}
\hline
& \multicolumn{2}{l}{Events per year} \\ \cline{2-3}
Oscillation scenario & \emph{IC-79} & \emph{IC3} \\ \hline
Normal ordering & 1.17 & 2.26 \\
Inverted ordering & 1.40  & 2.70 \\
\hline
\end{tabular}
\end{center}
\label{tab:icevents}
\end{table}

\begin{figure}%
\centering
\includegraphics[width=0.49\textwidth]{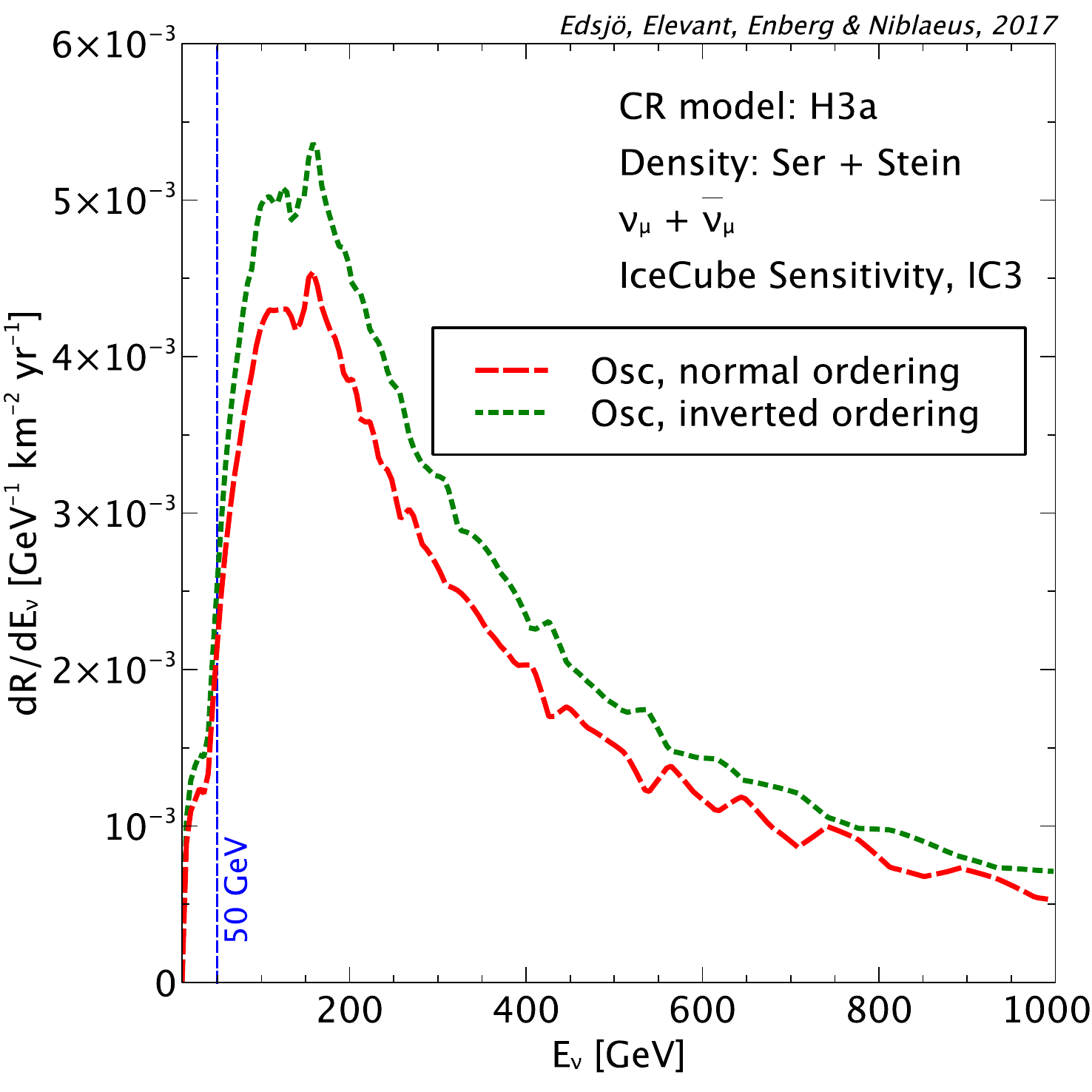}
\caption{The differential event rate $dR/dE_\nu$ for normal and inverted neutrino mass ordering for the effective areas of \emph{IC3}. In this figure we have included lower energies than our usual 50 GeV limit even if these are more uncertain.}
\label{fig:drde}
\end{figure}

In our study we have focused on neutrinos with energies larger than 50 GeV\@. As the differential spectrum falls steeply with energy one might wonder how sensitive the event rates in table \ref{tab:icevents} are to this lower energy limit. To test this, we can look at the differential rate, which for \emph{IC3} can be written as
\begin{equation}
\frac{dR}{dE_\nu} = \left( \frac{d\Phi_{\nu_\mu}}{dE_{\nu}} + \frac{d\Phi_{\bar{\nu}_\mu}}{dE_{\nu}} \right)
A_{\nu_\mu + \bar{\nu}_\mu}
\end{equation}
In figure \ref{fig:drde} we show this as a function of energy. We can see that our event rate is dominated by neutrinos of energies in the 100--300 GeV range and hence our calculated event rates are not very sensitive to the lowest energy of 50 GeV. This can be understood from the effective area, which is very small at low energies and then rises steeply as the energy goes up. In this interplay between the steeply falling neutrino spectrum and the increasing effective area we get a peak, which in this case happens to be in the 100--300 GeV region. The feature at around 40 GeV comes from the switch from the DeepCore to the IceCube selection in \emph{IC3} and the kink at 140 GeV is due to a kink in the IceCube effective area, whereas the wiggles at higher energies are due to oscillations.

\section{Comparisons with recent studies}

Just before finishing our work, two other studies focusing on similar aspects appeared \cite{Arguelles:2017eao,Ng:2017aur}.  The first one, by
Arg{\"u}elles et al. \cite{Arguelles:2017eao}, performs a very similar study as ours but in a slightly different framework. We have compared our results with their study, but have been informed that they are revising their calculation and paper. We therefore refrain from comparing with their results here.

The other study that appeared very recently is the study by Ng et al. \cite{Ng:2017aur}, hereafter NBPR. They do not recalculate the neutrino fluxes from cosmic ray interactions in the Sun but instead rely on earlier studies and focus on the neutrino floor this gives rise to for dark matter searches using neutrinos from the Sun. This aspect is also studied in ref.~\cite{Arguelles:2017eao}. There are many ways to present the fact that the \sanu flux presents a background for dark matter searches from the Sun and in 	NBPR they show results as a limit on the spin-dependent (SD) scattering cross section $\sigma_p^{SD}$ for different WIMP masses. The limits can be presented in this way by calculating the capture of WIMPs in the Sun and assume equilibrium between capture and annihilation (see e.g.\ ref. \cite{Wikstrom:2009kw}). There are of course uncertainties coming from both this assumption and from astrophysical input, most notably the local dark matter halo density and the velocity distribution. It is however a convenient way to present results as it makes it easy to compare with direct detection experiments directly sensitive to the scattering cross section. 

\begin{figure}%
\centering
\includegraphics[width=0.32\textwidth]{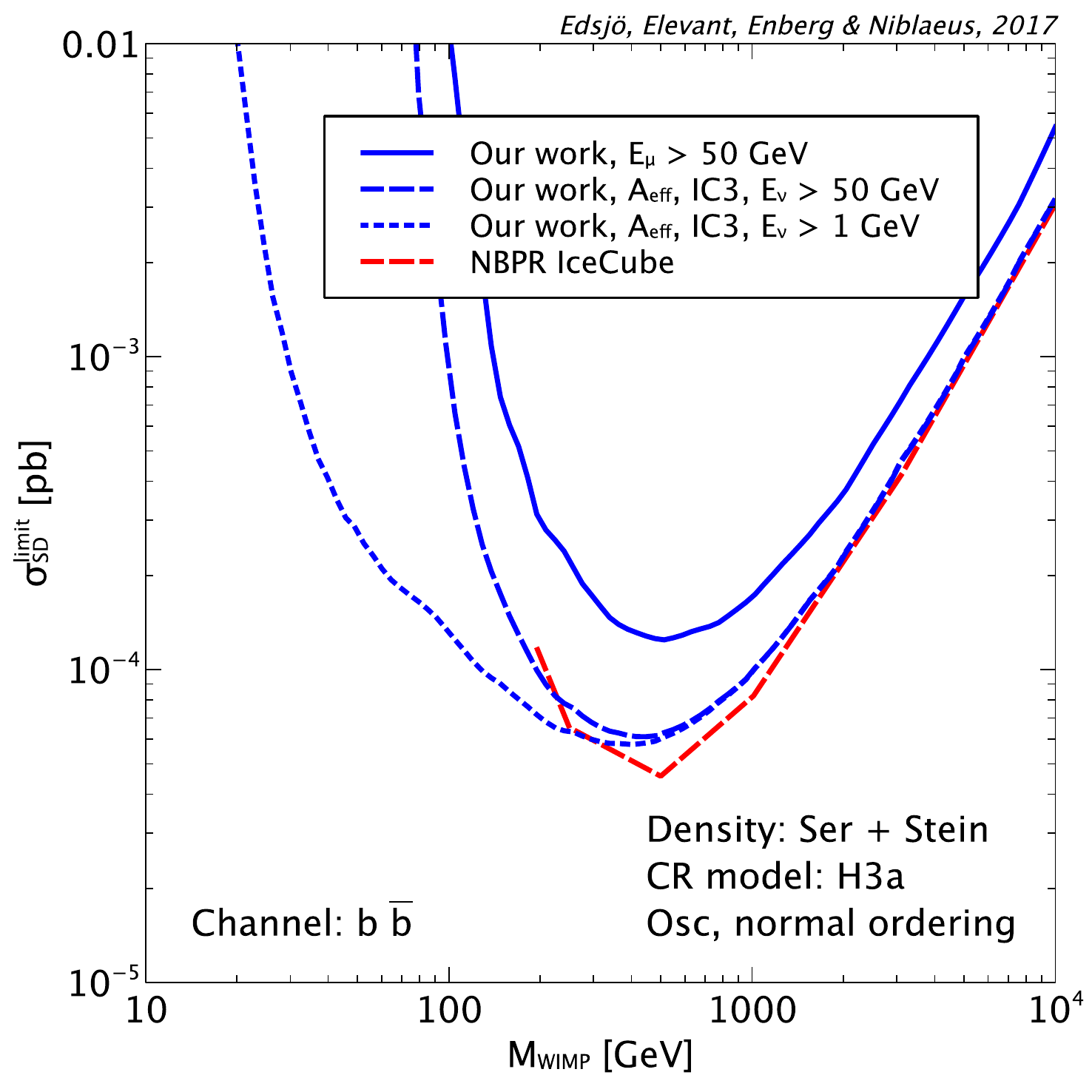}
\includegraphics[width=0.32\textwidth]{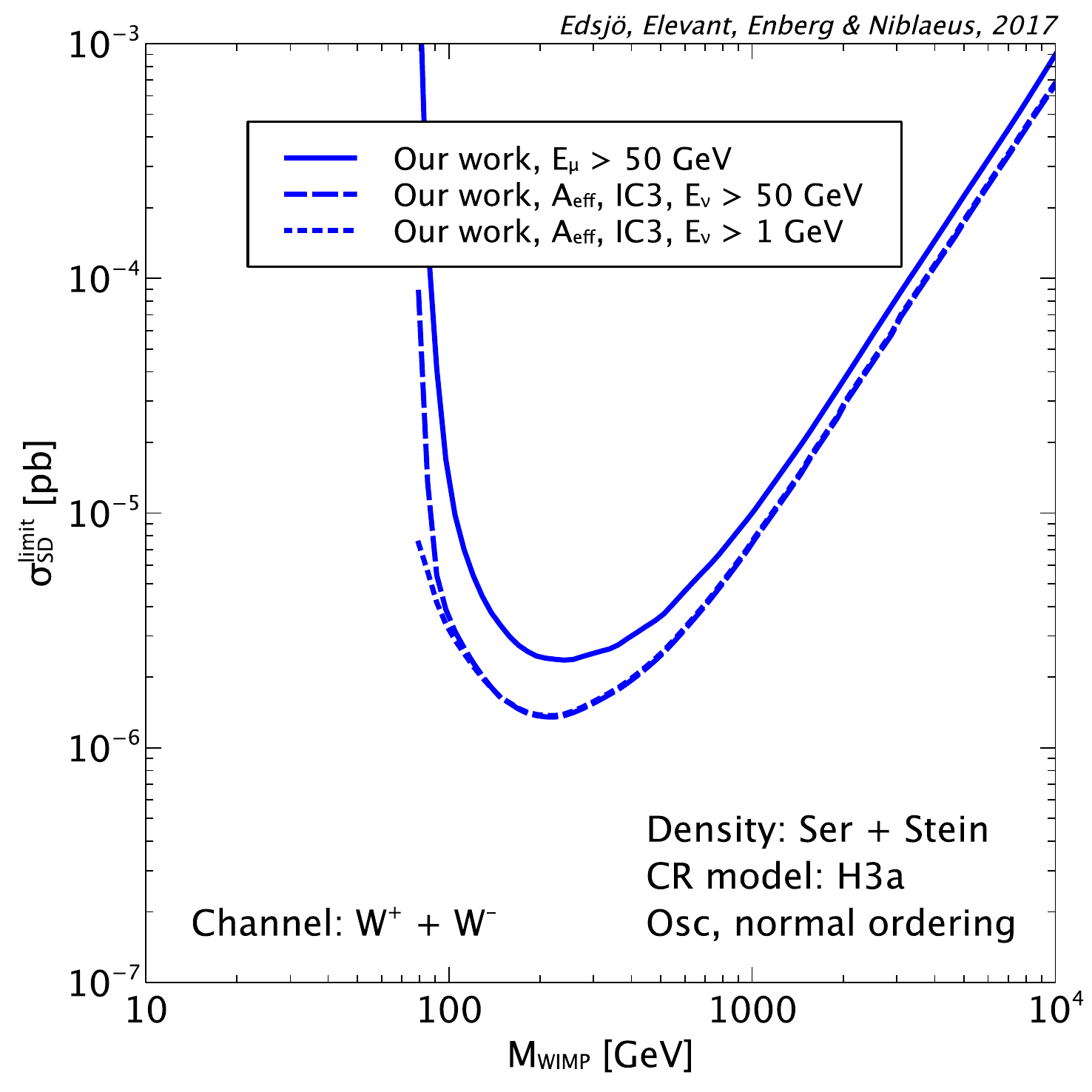}
\includegraphics[width=0.32\textwidth]{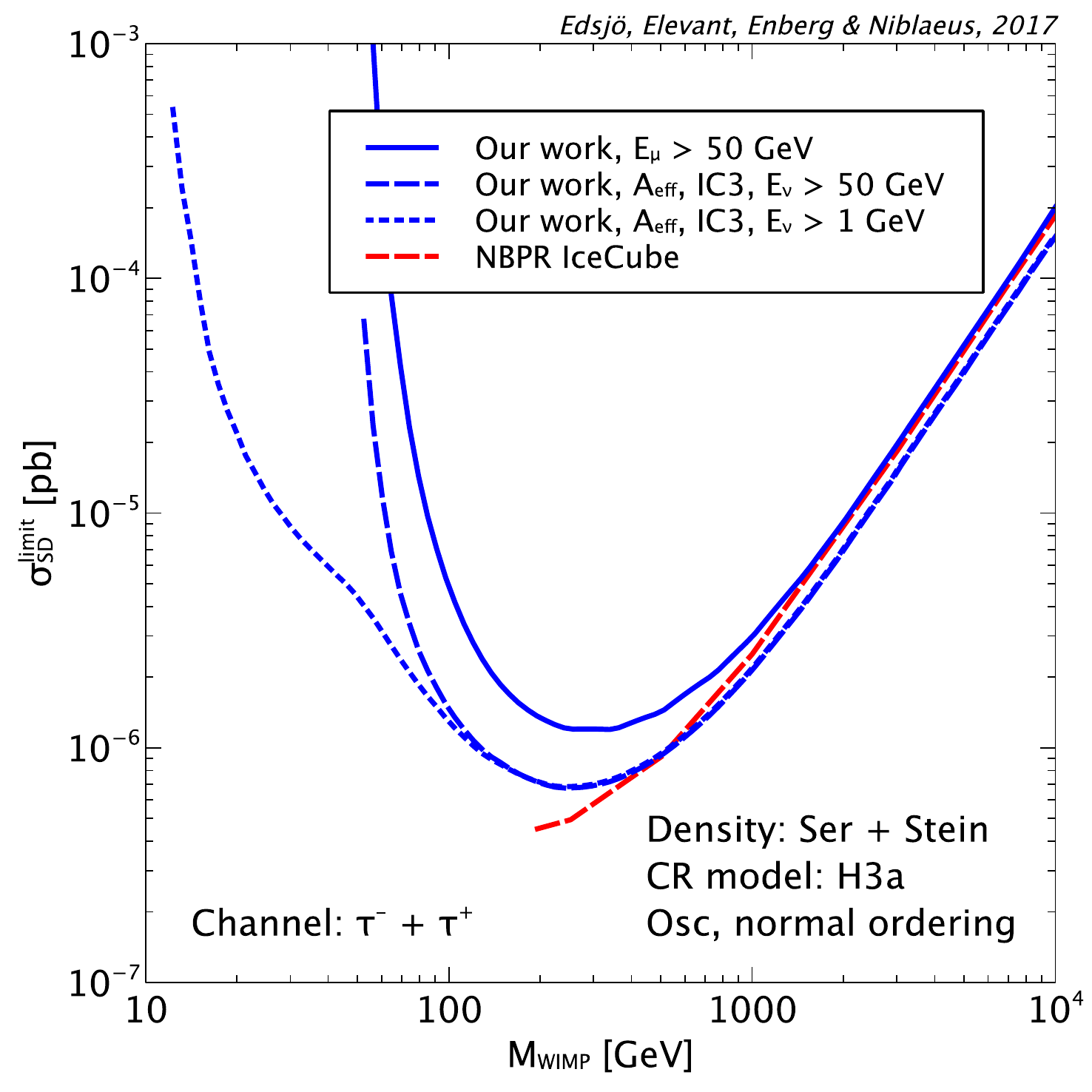}
\caption{The \sanu sensitivity floor expressed as a limit on the spin-dependent scattering cross section $\sigma_p^{SD}$ as a function of the WIMP mass. We show the sensitivity floor for $b\bar{b}$ (left), $W^+ W^-$ (middle) and $\tau^- \tau^+$ (right). For all three cases we show a simple estimate using a hard cut on the muon energy as a solid blue line, our estimate using the IceCube \emph{IC3} effective neutrino area integrating neutrino energies above $E_\nu=50$ GeV (dashed) and $E_\nu=1$ GeV (dotted). As our fluxes are more uncertain below 50 GeV, our primary results are the dashed blue curves. We also compare with the NBPR study (dashed red) \cite{Ng:2017aur}, where we use their naive curves. All results here are for IceCube.}
\label{fig:nufloor}
\end{figure}

We have made an estimate of this sensitivity floor using our results. In these estimates we calculate the rate in a neutrino telescope, where we use IceCube as an example. We calculate the capture and annihilation of WIMPs in the Sun with \darksusy \cite{Gondolo:2004sc} which in turn uses results from \wimpsim \cite{Edsjo:2007ws} for the WIMP annihilation signals. We also define the neutrino floor as the scattering cross section $\sigma_p^{SD}$ where we get equally many events from WIMP annihilations as from the \sanu flux.  We have calculated this estimate in three different ways: i) the first is to just look at the neutrino-induced muon fluxes (i.e.\ the ones in figure~\ref{fig:muoscfluxes}) and put a hard cut on the muon energy at 50 GeV, ii) the second is to instead use the neutrino fluxes directly and use the neutrino effective areas derived for dark matter searches, as done in section \ref{sec:events}, where we will here use the \emph{IC3} areas and only for $E_\nu>50$ GeV, and iii) the third way is the same as ii) but for $E_\nu>1$ GeV. Note that as we have large uncertainties below 50 GeV, the second choice is most robust for our calculation. The first method is similar to the method used in NBPR although they have also included contained events.

In figure~\ref{fig:nufloor} we show our results and for comparison also show the results (for IceCube) from the NBPR study. We note that our results agree fairly well with NBPR. That the NBPR $\tau^- \tau^+$ results do not show the turnover of the curve at around 200--300 GeV is most likely due to that they also include contained events.  To get a more accurate estimate of this neutrino  sensitivity floor one should really perform a detector simulation optimizing cuts (in e.g.\ energy and angle) to  reduce the \sanu background as much as possible. One should also perform a proper statistical analysis on what limits one could really set on WIMP dark matter given this partly irreducible \sanu background.

\section{Conclusion and outlook}

We have performed a new calculation of the solar atmospheric neutrino flux, \sanu. We also provide an event-based Monte Carlo package that can be used by experimental groups wishing to simulate this neutrino flux in their detector. Compared to earlier studies, we include both cosmic ray interactions, neutrino interactions and oscillations in a consistent framework. Even if our results qualitatively are quite similar to e.g. the earlier IT96 results \cite{Ingelman:1996mj}, our muon neutrino fluxes are lower, since neutrino oscillations reduce these fluxes. We also get lower fluxes at high energy which most likely comes from two effects, both that our production model gives less high energy neutrinos, and that our attenuation from interactions in the Sun is stronger than in the IT96 study. 

We have also compared our \sanu background to the signal from WIMP annihilation in the Sun and concluded that the energy and angular distribution of the two signals will be different, but not different enough to be able to discriminate the two effectively. Hence, the \sanu flux will be an essentially irreducible background for neutrino searches from WIMP annihilation in the Sun. To properly address how significant this background is will be tough, requiring a detailed detector simulation to also include detector effects like reconstruction uncertainties on the muon energy and angle.

We have also compared to Earth's atmospheric neutrino background and conclude that the \sanu flux is higher in the direction of the Sun, especially at higher energies. The total event rates are rather low though with only a few events per year even in a large detector, like IceCube.

We also point out some future directions. We have in this study used the \mceq simulation package which only include all parts of the neutrino production flux above 50 GeV. Although the rates in current neutrino telescopes are mostly sensitive to \sanu at higher energies,
it would be very interesting to investigate lower energies in a future study,
as thresholds of neutrino telescope are lower than this. However, in that case one would also need to address the difficult question of the impact of the magnetic fields in the Solar System that will effect the lower energies. One possible way forward in this regard could be to correlate the neutrino flux with the gamma ray flux measured by the Fermi-LAT and e.g.\ IceCube. 

As the gamma ray flux would depend on the magnetic fields in similar ways, it should be possible to reduce the uncertainties this way. It would also be worthwhile to investigate more systematically different cosmic ray models and hadronic interaction models to get a better handle on how large the uncertainties on the predicted \sanu fluxes are.

\section*{Acknowledgements}
We thank A. Fedynitch for help with \mceq in general, and for providing an update and help with proton-proton tables in particular. We also thank Robert Stein for help using his MHD simulation data. We also thank the anonymous referee for very helpful comments.


\appendix

\section{The WimpSim code}
\label{app:wimpsim}

\begin{figure}
\centering
\includegraphics[width=\textwidth]{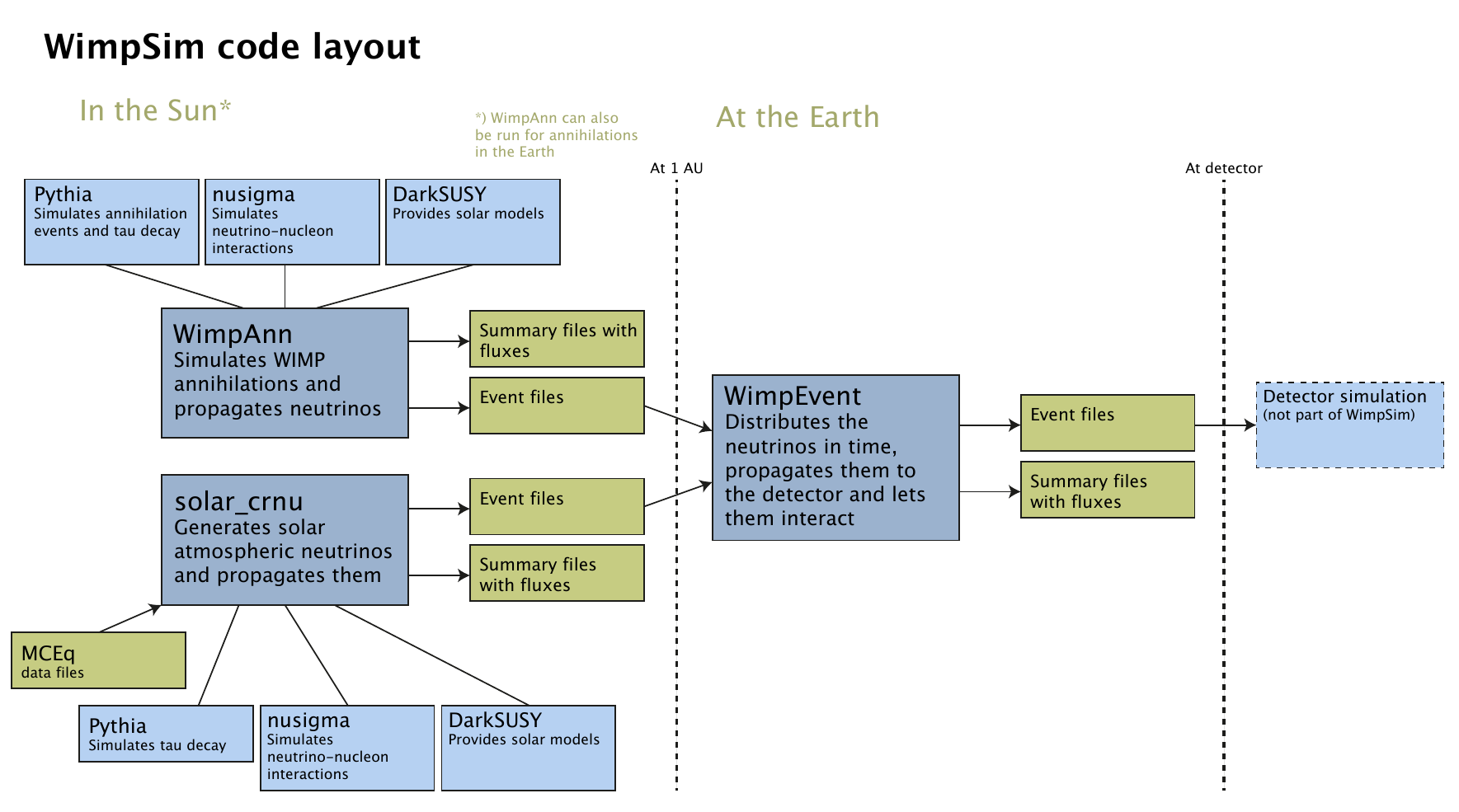}
\caption{The layout of the WimpSim code. Solid lines indicate linked codes and arrows indicate reading/writing of data files.}
\label{fig:wimpsim}
\end{figure}

The \wimpsim package contains three main codes: \wimpann, \solarcrnu and \wimpevent, see figure~\ref{fig:wimpsim} for an illustration of the layout of the code. \wimpann and \wimpevent are used to simulate dark matter WIMP annihilations in the Sun (and the Earth) and its production of neutrinos and their interactions and oscillations. Interactions are calculated with our neutrino-nucleon Monte Carlo \nusigma. \solarcrnu is our new add-on that reads the output from our modified \mceq and generates solar atmospheric neutrinos and lets these interact and oscillate. In practice, \solarcrnu will take the neutrinos from their starting point and let them interact and oscillate through the Sun. It will then take them to 1 AU from the Sun (Earth's average distance from the Sun). At this point, both summary and event files will be saved. The summary files contain the fluxes of neutrinos at production, at the solar surface and at 1 AU, whereas the event files contain the actual neutrino events together with their energy, weight and state vector (amplitude and phases in the neutrino interaction base). 

\wimpevent will take these event files as input and propagate the neutrinos further to our actual detector where it will let them interact (again using \nusigma as our neutrino-nucleon simulation software) and produce leptons and hadronic showers. We are here mostly interested in the muons coming from charged interactions, but the code can simulate both neutral and charged currents to be used for neutrino telescope simulations. In case a muon is produced, we also let it propagate in the medium surrounding the detector where it will undergo energy losses and multiple Coulomb scattering. \wimpevent will also produce both summary and event files. The summary files contain fluxes of neutrinos, charged leptons at the neutrino-nucleon interaction point, muons after propagation to the detector and hadronic showers at the interaction point, whereas the event files contain both the incoming neutrino and the lepton and hadronic shower at the neutrino-nucleon interaction point. These event files are suitable for further detector simulation. \wimpevent will take the actual detector location into account and will perform the simulation over a given time frame and properly time stamp the events and include both particle directions relative to the Sun and in usual astronomical coordinates. For this we use \texttt{SLALIB} \cite{Wallace:2012slalib}.

The main modifications we have made to the code to include the \sanu{}s are to add reading of \mceq output files and drawing events from those distributions. We have also modified the code to allow for arbitrary neutrino paths through the Sun (earlier only radial parts were included). Compared to the earlier publication describing \wimpsim \cite{Blennow:2007tw}, there have also been other improvements to the code that are not directly related to the \sanu{}s. E.g.\ the code now uses a much more accurate time stamp of the events giving modified Julian dates (MJD) for the events with proper directions of all the produced particles. 

We have optimised the code for speed without sacrificing accuracy. For example, for the event generation from \mceq production fluxes, we gain speed by using a good test function for our acceptance-rejection sampling. The choice of test function does not affect the accuracy, only the speed. For the propagation and oscillation simulation, we have optimised the step size to use reasonably sized steps without sacrificing accuracy (where we require accuracy at the percent level). Typical timings for a simulation run with 1 million neutrino events (on an Intel i7 2.6 GHz CPU ) are that about 1 minute is spent on drawing events from the \mceq production fluxes, about 9 minutes are spent on neutrino propagation and oscillations in \wimpann and finally about 1 minute and 20 seconds are spent in \wimpevent propagating the neutrinos to a detector and simulating interactions at the detector. So, in total about 11.3 minutes per 1 million events. In this paper, we have generated 250 million events for 12 scenarios so we have spent about 25 CPU days to get our results. 

For details we refer the interested reader to refs.\ \cite{Blennow:2007tw,Edsjo:2007ws}. Our updated codes (both \wimpsim and \nusigma) are available at \cite{Edsjo:2017ws} and we refer the interested reader to this code web page for more details. 


\bibliographystyle{JHEP}
\bibliography{solar}

\end{document}